\documentclass[%
nofootinbib,
amsmath,amssymb,
aps, reprint
]{revtex4-2}
\usepackage{hyperref} 
\usepackage{graphicx}
\usepackage{dcolumn}
\usepackage{bm}
\usepackage{graphicx}
\usepackage{epstopdf}
\usepackage{mathrsfs}
\usepackage{amssymb}
\usepackage{slashed}
\usepackage{verbatim}
\usepackage{color}
\usepackage{multirow}
\usepackage{amsmath}
\usepackage{epsfig}
\usepackage{ulem}
\usepackage{graphics}
\usepackage{indentfirst}
\usepackage{slashed}
\usepackage{amscd}
\usepackage[all,cmtip]{xy}
\usepackage{youngtab}

\usepackage{simplewick}
\usepackage{changepage}
\usepackage{bm}

\newcommand{\bea}{\begin{eqnarray}}
\newcommand{\eea}{\end{eqnarray}}
\newcommand{\bes}{\begin{align*}}
\newcommand{\ees}{\end{align*}}

\newcommand\nn{\nonumber}

\newcommand{\ad}{\dot{\alpha}}
\newcommand{\bd}{\dot{\beta}}
\newcommand{\rd}{\dot{\rho}}
\newcommand{\sid}{\dot{\sigma}}

\def\beq{\begin{equation}}
\def\eeq{\end{equation}}
\def\beqs{\begin{equation*}}
\def\eeqs{\end{equation*}}
\def\<{\langle}
\def\>{\rangle}

\def\nn{\nonumber}

\def\cO {{\cal O}}
\def\cG {{\cal G}}

\newcommand{\cT}{\mathcal{T}}

\newcommand{\mT}{\mathscr{T}}
\newcommand{\mG}{\mathscr{G}}

\newcommand{\cM}{\mathcal{M}}

\def\cR{{\cal R}}
 
\begin{document}

\begin{abstract}
A program of wide interest in modern conformal bootstrap studies is to numerically solve general conformal field theories, based on a critical assumption that the dynamics is encoded in the conformal four-point crossing equations and positivity condition. 
In this letter we propose and verify a novel algebraic property of the  crossing equations which provides strong restriction for this program. We show for various types of symmetries $\cal G$, the crossing equations can be linearly converted into the $SO(N)$ vector crossing equations associated with the $SO(N)\rightarrow \cal G$ branching rules and the transformations satisfy positivity condition. The  dynamics constrained by the $\cal G$-symmetric crossing equations combined with positivity condition degenerates to the $SO(N)$ symmetric cases, while the non-$SO(N)$ symmetric theories are not directly solvable  without introducing the $SO(N)$ symmetry breaking assumptions on the spectrum.

\end{abstract}

\title{Symmetries of conformal correlation functions}

\author{ Zhijin Li}

\affiliation{
Department of Physics, Yale University, New Haven, CT 06511}

\maketitle

\section{Introduction}
The bootstrap approach has been proposed in the early stage of quantum field theory studies, which aims to solve theories using general consistency criteria, particularly the crossing symmetry and unitarity. The bootstrap method  for conformal field theories (CFTs) has been proposed in \cite{Polyakov:1974gs, Ferrara:1973yt} with remarkable successes in the classification of 2D rational CFTs \cite{Belavin:1984vu} and the reviving of modern conformal bootstrap \cite{Rattazzi:2008pe, Poland:2018epd}, see e.g. \cite{El-Showk:2012cjh, El-Showk:2014dwa, Kos:2014bka, Kos:2015mba} for part of the significant developments. Stimulated by the success of modern conformal bootstrap, it is widely interested whether general CFTs can be solved using bootstrap approach, as also emphasised in \cite{Poland:2018epd}: ``{\it can we use the bootstrap to fully classify the space of critical CFTs with a given symmetry...?}".
For a majority of CFTs with global symmetries, this project can be rephrased as: {\it can the CFTs with a symmetry $\cG$ be solved using the $\cG$-symmetric crossing equations combined with positivity and few general assumptions on the spectrum?}
The key assumption is that the dynamics of general CFTs is encoded in the crossing equations and positivity condition.
In this work, we will propose and verify a novel algebraic property that the conformal four-point crossing equations with different global symmetries are actually endowed with an $SO(N)$ symmetric positive structure, which leads to strong challenges for the above project.


The motivation of this work originates from rapid developments in conformal bootstrap studies    revived since the breakthrough work \cite{Rattazzi:2008pe}. Among several remarkable developments \cite{El-Showk:2012cjh, El-Showk:2014dwa, Kos:2014bka, Kos:2015mba},  there is a mysterious phenomenon: the bootstrap bounds from crossing equations with different global symmetries can coincide with each other although the bootstrap equations have distinct forms for different symmetries. Such  bound coincidences were firstly observed in \cite{Poland:2011ey}, which shows the singlet bounds from the $SO(2N)$ vector and $SU(N)$ fundamental bootstraps are the same. More examples for the bootstrap bound coincidences with  different symmetries have been observed in \cite{Caracciolo:2014cxa, Nakayama:2016knq,  Nakayama:2017vdd, Li:2018lyb, Stergiou:2018gjj, Stergiou:2019dcv, Li:2020bnb}. 
The bound coincidences indicate intriguing connections between the bootstrap algorithm and symmetry properties of the crossing equations. This puzzle has been solved recently in \cite{Li:2020bnb}, which discovered a subtle algebraic relation between the $SU(N)$ fundamental and $SO(2N)$ vector crossing equations: the two sets of crossing equations are connected by a linear transformation consistent with the positivity constraint in the bootstrap algorithm. Due to this relation, the linear functionals for the $SO(2N)$ vector bootstrap can be used to construct the linear functionals for the $SU(N)$ fundamental bootstrap, which further leads to the bootstrap bound coincidences.
In this work, we will provide convincing evidence for similar algebraic relations of general global symmetries. We will show that this algebraic structure has surprising applications on the symmetries of conformal four-point correlation functions and the conformal bootstrap program.

\section{Symmetries in crossing equations}
Crossing equations are obtained by taking operator product expansions (OPEs) of the conformal four-point correlator $\langle \cO_1\cO_2\cO_3\cO_4\rangle$ in either $(\cO_1\cO_2)$-$(\cO_3\cO_4)$ channel (s-channel) or $(\cO_1\cO_4)$-$(\cO_2\cO_3)$ channel (t-channel), which lead to different conformal partial wave (CPW) expansions of the same correlator.  They introduce dynamical constraints for the bootstrap computations. For CFTs with a global symmetry $\cG$, the crossing equations possess an algebraic structure related to the $6j$ symbols of the group $\cG$. 
A systematical approach for crossing equations with global symmetries has been provided  in \cite{Rattazzi:2010yc}.\footnote{The crossing equations strongly depend on the symmetries and one has to compute the crossing equations case by case. In \cite{Go:2019lke} a {\it Mathematica} program has been developed to compute the crossing equations with global symmetries automatically.} 
Considering a scalar $\cO_\alpha$ which constructs an irreducible (complex) representation $\pi$ of the group $\cG$, its four-point correlator can be expanded as 
 {\small \bea 
\langle \cO_\alpha (x_1)\cO^\dagger_{\bar{\alpha}}(x_2) \cO_\beta(x_3) \cO^\dagger_{\bar{\beta}}(x_4) \rangle&=& x_{12}^{-2\Delta_\cO}x_{34}^{-2\Delta_\cO} \mG_\cO(u,v),  \nn   \\
&&\hspace{-4.3cm}
\mG_\cO(u,v)=\sum_{i,j} \cT^{R_i}_{\alpha \bar{\alpha} \beta \bar{\beta}} G^+_{R_i}(u,v)+ \cT^{r_j}_{\alpha \bar{\alpha} \beta \bar{\beta}} G^-_{r_j}(u,v),~~ \label{schannel}
\eea }
where  $\cT^{\pi_i}$ are the $\cG$-invariant  four-point  tensors  obtained by contracting two Clebsch-Gordan (CG) coefficients $(\pi\otimes \pi^\dagger\otimes \pi_i)$ and $({\pi_i}^\dagger\otimes\pi\otimes \pi^\dagger)$. Superscripts in
$G_{\pi_i}^\pm$ denote the correlation functions even/odd under the interchange of external coordinates $x_1\leftrightarrow x_2$ or $x_3\leftrightarrow x_4$. In the CPW expansions, they respectively relate to even/odd spin selection rules.\footnote{Under the interchange of $x_1\leftrightarrow x_2$ or $x_3\leftrightarrow x_4$, the conformal blocks appearing in the CPW expansion of (\ref{schannel}) only acquire a spin $\ell$-dependent factor $(-1)^\ell$ \cite{Dolan:2000ut, Dolan:2011dv}, therefore only even(odd) spins appear in the CPW expansion of $G_X^+(G_X^-)$. } Variables $u,v$ are the conformal invariant cross ratios 
$
u\equiv  x_{12}^2x_{34}^2/x_{13}^2x_{24}^2,  v\equiv  x_{14}^2x_{23}^2/x_{13}^2x_{24}^2
$. 
This correlator can be alternatively expanded in the t-channel by exchanging $\{x_2, \bar{\alpha}\} \leftrightarrow \{x_4, \bar{\beta}\} $ in (\ref{schannel}). The two expansions lead to a tensor equation
{\small
\bea
v^{\Delta_\cO}\sum_{i,j} \cT^{R_i}_{\alpha \bar{\alpha} \beta \bar{\beta}} G^+_{R_i}(u,v)+ \cT^{r_j}_{\alpha \bar{\alpha} \beta \bar{\beta}} G^-_{r_j}(u,v) &=& \label{crq1}\\ &&\hspace{-5cm} u^{\Delta_\cO}
\sum_{i,j} \cT^{R_i}_{\alpha \bar{\beta} \beta \bar{\alpha}} G^+_{R_i}(v,u)+ \cT^{r_j}_{\alpha \bar{\beta} \beta \bar{\alpha}} G^-_{r_j}(v,u).
\nn
\eea}
Constraints from crossing symmetry are completed by considering another configuration of  the four-point correlator  $\langle \cO_\alpha (x_1)\cO^\dagger_{\bar{\alpha}}(x_2)\cO^\dagger_{\bar{\beta}}(x_3)  \cO_\beta(x_4)\rangle$.\footnote{The equation (\ref{crq1}) provides sufficient constraints if the representation $\cR$ is real.}   
Crossing equations can be obtained by decomposing the $\cG$-invariant four-point tensors in one channel to another channel, which is given by the inner products of s-channel and t-channel $\cG$-invariant four-point tensors, namely the $6j$ symbols of group $\cG$.
In conformal bootstrap studies, the correlation functions $G_{\pi_i}^\pm$ are further decomposed into infinite sums of conformal blocks \cite{Dolan:2000ut,Dolan:2003hv, Dolan:2011dv}, which are reminiscent to the $\cG$-invariant  four-point tensors but with CG coefficients replaced by the conformal three-point correlation functions, and their crossing symmetric decomposition is given by the $6j$ symbols of the conformal group \cite{Gadde:2017sjg, Liu:2018jhs}.

The $SO(N)$ vector  four-point correlator $\langle \phi_i\phi_j\phi_k\phi_l\rangle$ contains three $SO(N)$-invariant four-point tensors {\small
\bea
\mG_\phi(u,v) = \delta_{ij}\delta_{kl}G_S^+(u,v)+\left(\delta_{il}\delta_{jk}-\delta_{ik}\delta_{jl}\right)G_A^-(u,v)  \label{cAphi} \\ 
  + \left(\delta_{ik}\delta_{jl}+\delta_{il}\delta_{jk}-2\delta_{ij}\delta_{kl}/N\right)G_T^+(u,v),\nn
\eea }
where $S/T/A$ denote $SO(N)$ singlet, traceless symmetric and anti-symmetric representations. 
Crossing symmetry requires the three functions $G_{S/T/A}(u,v)$ satisfying \cite{Rattazzi:2010yc}:
\bea
\left(
\begin{array}{ccc}
 0 & F_T^+ & -F_A^- \\
 F_S^+ & \left(1-\frac{2}{N}\right) F_T^+ & F_A^- \\
 H_S^+ & -\left(1+\frac{2}{N}\right) H_T^+ & -H_A^- \\
\end{array}
\right)\cdot
\left(
\begin{array}{c}
 1 \\
 1 \\
 1 \\
\end{array}
\right)=\left(
\begin{array}{c}
 0 \\
 0 \\
 0 \\
\end{array}
\right), \label{crqON}
\eea
where the functions $H_{\pi_i}^\pm(F_{\pi_i}^\pm)$ are (anti-)symmetric linear superposition of $G^\pm_{\pi_i}(u,v)$ and $G^\pm_{\pi_i}(v,u)$
\bea
F^\pm_{\pi_i}(u,v) &\equiv &v^{\Delta_{\phi}}G^\pm_{\pi_i}(u,v)-u^{\Delta_\phi}G^\pm_{\pi_i}(v,u),  \\
H^\pm_{\pi_i}(u,v)&\equiv &v^{\Delta_\phi}G^\pm_{\pi_i}(u,v)+u^{\Delta_\phi}G^\pm_{\pi_i}(v,u). 
\eea
Surprisingly, for a large class of examples in previous bootstrap studies \cite{Poland:2018epd}, we find the crossing equations can be linearly transformed into the $SO(N)$ vector's (\ref{crqON}), although they have rather distinct forms.
Let us take the four-point crossing equations of $SU(N)$ fundamental scalars as an example. The $SU(N)$ fundamental crossing equations can be written in a matrix form \cite{Rattazzi:2010yc}:
\begin{widetext}
\bea
\cM_{SU(N)}\cdot \mathbf{1}_{6\times1}\equiv\left(
\begin{array}{cccccc}
 0 & 0 & F_{\text{Adj}}^+ & -F_{\text{Adj}}^- & F_T^+ & -F_A^- \\
 0 & 0 & H_{\text{Adj}}^+ & -H_{\text{Adj}}^- & -H_T^+ & H_T^- \\
 F_S^+ & F_S^- & \left(1-\frac{1}{N}\right) F_{\text{Adj}}^+ & \left(1-\frac{1}{N}\right) F_{\text{Adj}}^- & 0 & 0 \\
 H_S^+ & H_S^- & -\left(1+\frac{1}{N}\right) H_{\text{Adj}}^+ & -\left(1+\frac{1}{N}\right) H_{\text{Adj}}^- & 0 & 0 \\
 F_S^+ & -F_S^- & -\frac{1}{N}F_{\text{Adj}}^+ & \frac{1}{N}F_{\text{Adj}}^- & F_T^+ & F_A^- \\
 H_S^+ & -H_S^- & -\frac{1}{N} H_{\text{Adj}}^+ & \frac{1}{N} H_{\text{Adj}}^- & -H_T^+ & -H_T^- \\
\end{array}
\right)\cdot 
\left(
\begin{array}{c}
 1 \\
 1 \\
 1 \\
 1 \\
 1 \\
 1 \\
\end{array}
\right)=
\left(
\begin{array}{c}
 0 \\
 0 \\
 0 \\
 0 \\
 0 \\
 0 \\
\end{array}
\right), \label{crqSU}
\eea 
\end{widetext}
where the $S$ (singlet) and Adj (adjoint) sectors appear in $\Phi \times\Phi^\dagger \sim S^\pm+\textrm{Adj}^\pm$ while the $T$ (symmetric) and $A$ (anti-symmetric) sectors appear in $\Phi\times \Phi \sim T^++A^-$. 
 
Crossing equations of the $SO(2N)$ vector scalars (\ref{crqON}) and $SU(N)$ fundamental scalars (\ref{crqSU}) have quite different forms. Remarkably, the two sets of crossing equations are actually related through a linear transformation $\mT_{SU(N)}$ given by the matrix \cite{Li:2020bnb}: 
\footnote{The transformation $\mG_{SU(N)}$ can be solved by imposing the branching rules (\ref{branching4}-\ref{branching6}) and another constraint that there is no mixing between the functions $F$ and $H$, which have opposite symmetries under  $u\leftrightarrow v$. These constraints lead to over-determined equations, which surprisingly can be solved uniquely up to normalization.}
{\small
\beq
\left(
\begin{array}{cccccc}
 1 & 0 & \frac{1}{1-2 N} & 0 & \frac{1}{2 N-1} & 0 \\
 0 & 0 & \frac{1}{2 N-1}+1 & 0 & \frac{1}{1-2 N}+1 & 0 \\
 0 & \frac{2}{2 N-1} & 0 & \frac{1}{1-2 N}+1 & 0 & \frac{1}{2 N-1}+1 \\
\end{array} 
\right). \label{Ts}
\eeq }
With action $\mT_{SU(N)}$, the equation (\ref{crqSU}) turns into
{\small
\begin{widetext}
 \bea
\mathscr{T}_{SU(N)}\cdot \cM_{SU(N)}\cdot \mathbf{1}_{6\times1} &=& \left(
\begin{array}{cccccc}
 0 &  -  F_S^- &  F_{\text{Adj}}^+&  - F_{\text{Adj}}^- &  F_T^+& -F_A^- \\
 F_S^+ &   F_S^- &  \left(1-\frac{1}{N}\right) F_{\text{Adj}}^+ &  F_{\text{Adj}}^- &  \left(1-\frac{1}{N}\right) F_T^+ &  F_A^- \\
 H_S^+ &  -H_S^- & -\left(1+\frac{1}{N}\right) H_{\text{Adj}}^+ & -H_{\text{Adj}}^- & -\left(1+\frac{1}{N}\right) H_T^+ & -H_A^- \\
\end{array}
\right) \cdot
\left(
\begin{array}{c}
 1 \\
 y_1 \\
 x_1 \\
 y_2 \\
 x_2 \\
 y_3 \\
\end{array}
\right)=
\left(
\begin{array}{c}
 0 \\
 0 \\
 0
\end{array}
\right), ~~~ \label{SU2SO}
\\
(y_1,~x_1,~y_2,~x_2,~y_3)&=& \left(\frac{1}{2 N-1},\frac{N-1}{2 N-1},\frac{N^2-1}{N(2 N-1)},\frac{N}{2 N-1},\frac{N-1}{2 N-1}\right).\label{xs} 
\eea
\end{widetext}} 
We call the constants $x_i,y_j$ ``{\it recombination coefficients}".
The $3\times6$ matrix in (\ref{SU2SO}) is equivalent to the $3\times 3$ matrix in the $SO(2N)$ vector crossing equations  (\ref{crqON}) combined with the $SO(2N)\rightarrow SU(N)$ branching rules {\small
\bea
SO(2N) & \hspace{2cm}  &  SU(N) \nn \\
S ~&\longrightarrow&  ~S,  \label{branching4}\\
T ~&\longrightarrow& ~  \text{Adj}  \oplus  T, \label{branching5}\\
A~ &\longrightarrow& ~ S\oplus  \text{Adj} \oplus A. \label{branching6}
\eea }
Besides, it is critical that the coefficients $x_i, y_j$ in (\ref{xs}) are all positive for $N>1$, which guarantees that the positivity in the CPW expansions will not be changed by the transformation. 

Now we define a new $SO(2N)$-symmetric correlation function $\mG_{\phi'}$ (\ref{cAphi}) of a presumed $SO(2N)$ vector $\phi'_i$ based on the $SU(N)$ fundamental correlation functions $G_{\pi_i}^\pm$ and the recombination coefficients $x_i,y_j$: {\small
\bea
{G}_S'^{\,+}(u,v) &=& G_S^+(u,v), \label{newGS} \\
G_T'^{\,+}(u,v) &=& x_1 G_ {\text{Adj}}^+(u,v) +x_2G_T^+(u,v) ,   \label{newGT}\\
G_A'^{\,-}(u,v)  &=& y_1G_S^-(u,v)+y_2 G_{\text{Adj}}^-(u,v) +y_3 G_A^-(u,v). ~~ \label{newGA}
\eea }
Because of the constraints  (\ref{SU2SO}), the above correlation functions $\mG_{\phi'}$  satisfy the $SO(2N)$ crossing equations  (\ref{crqON})! Moreover, 
since the recombination coefficients $x_i, y_j$ are positive, as long as the original $SU(N)$ correlator admits CPW expansions with unitary coefficients, so does the new $SO(2N)$ symmetric correlator. Such construction will be dubbed $SO(N)$-ization.\footnote{We thank David Poland and Slava Rychkov for discussions and suggestions on the terminology.} We present some examples for $SO(N)$-ization in Section \ref{App-GFTs} and \ref{app:WZW}. 

We have obtained linear transformations  $\mT_{\cG}$ similar to $\mT_{SU(N)}$ in (\ref{SU2SO}) for plenty of symmetry groups appeared in previous bootstrap studies, including the discrete symmetries, classical and exceptional Lie groups with fundamental or higher rank representations, see a review in \cite{Poland:2018epd} for related bootstrap studies. Part of the interesting examples are presented in Section \ref{app:moregps}.  
Due to the transformation $\mT_{\cG}$, the $SO(N)$ symmetric correlation function $\mG_{\phi'}$ can be constructed based on the original $\cG$-symmetric correlation functions $\mG_{\cO}$ similar to (\ref{newGS}-\ref{newGA}). 
The diversified examples suggest the linear transformation $\mT_{\cG}$ and $SO(N)$-ization are true for general $\cG$, which can be summarized as follows:
\newline
\hspace*{0.5cm}Consider the four-point correlator of a scalar $\cO$ with $N$ real components which forms an irreducible representation $\pi$  of a group $\cG$. The correlation function consists of components $G_{\pi_i}^\pm$ multiplied with invariant four-point tensors of $\cG$ irreducible representations $\pi_i$, which can be classified into three parts: the $\cG$ singlet $S_\cG$, and the $\cG$ non-singlet representations $R_i (r_j)$ which are parity even (odd) under the interchange of external coordinates $x_1\leftrightarrow x_2$. The components $G_{\pi_i}^\pm$ satisfy the $\cG$-symmetric crossing equations $\cM_{\cG}$.
We conjecture: {\bf a,} there is a linear transformation (unique up to normalization)  $\mT_{\cG}$,  which maps the crossing equations $\cM_{\cG}$ to the $SO(N)$-symmetric crossing equations (\ref{crqON}) \footnote{It has been proved in \cite{Rattazzi:2010yc} that the single four-point correlator crossing equations $\cM_\cG$ can always be given by an $n\times n$ square matrix}
{\small
\bea
\mT_{\cG_{3\times n}}\cdot \cM_{\cG_{n\times n}}\cdot \mathbf{1}_{n\times1} &=& \label{G2SO} \\  && \hspace{-3.5cm}
 \left(
\begin{array}{ccccc}
 0 &  F_{R_i}^+ &  \dots&   -F_{r_j}^- &  \dots  \\
 F_S^+ &   \left(1-\frac{2}{N}\right)F_{R_i}^+ &  \dots &  F_{r_j}^- &  \dots \\
 H_S^+ &  -\left(1+\frac{2}{N}\right)H_{R_i}^+ & \dots & -H_{r_j}^- & \dots \\
\end{array}
\right) \cdot
\left(
\begin{array}{c}
 1 \\
 x_{R_i} \\
 \vdots \\
 y_{r_j} \\
 \vdots \\
\end{array}
\right)=\mathbf{0}_{3\times1}  \nn 
\eea }
associated with the
 $SO(N)\rightarrow \cG$ branching rules
\beq
S_{SO(N)} \rightarrow S_{\cG}, ~ 
T_{SO(N)} \rightarrow  \oplus_i  R_i, ~
A_{SO(N)} \rightarrow  \oplus_j r_j,\label{gbrchso}
\eeq
and {\bf b,} all the recombination coefficients $x_{i}, y_{j}$ are positive.
Consequently, a new $SO(N)$-symmetric correlation function $\mG_{\phi'}$ can be constructed from  correlation functions $G_{\pi_i}^\pm$ and the recombination coefficients $x_{i}, y_{j}$: {\small
{\small
\bea
{G}_S'^{\,+}(u,v) &=& G_S^+(u,v),  ~~
G_T'^{\,+}(u,v) = \sum_{i} x_{i} G_ {R_i}^+(u,v),  ~~~ \label{gnewGS}\\
G_A'^{\,-}(u,v)  &=& \sum_{j} y_{j}G_{r_j}^-(u,v).  \label{gnewGA}
\eea }
This is strongly against our intuition about symmetries in physics: generically it requires restrictive conditions to have enhanced symmetries in physical systems, while above proposal suggests that due to a dedicate positive structure in the crossing equations, the conformal four-point correlation function can  be linearly transformed into a form with a maximal symmetry allowed by its degree of freedom!
\footnote{The $SO(N)$-ized correlators in (\ref{gnewGS}-\ref{gnewGA}) are not necessarily related to locally interacting theories. A counter example is provided by the $SO(N)$-ization of the 3D cubic model with $\mathbb{Z}_2^N\rtimes S_N$ symmetry, see \ref{app:cubic}. The theory does not have a spin 1 conserved current, nor does its $SO(N)$-ized correlator. Therefore the later is not of  $SO(N)$ symmetric theories with local interactions. }

\section{Symmetries in conformal bootstrap}
The algebraic relations (\ref{G2SO},\ref{gbrchso}) associated with positive recombination coefficients have substantial applications in conformal bootstrap studies. To bootstrap the four-point correlator $\langle \cO\cO^\dagger\cO\cO^\dagger\rangle$, one takes the conformal block expansions of the correlation functions. The bootstrap crossing equations have the same matrix form as $\cM_{\cG}$ but with the correlation functions $G_{\pi_i}^\pm$ replaced by conformal blocks. The bootstrap algorithm \cite{Rychkov:2016iqz, Simmons-Duffin:2016gjk,Chester:2019wfx} aims to find $n$-component linear functions $\vec{\alpha}_{\cG}$ whose actions on $\cM_{\cG}$ satisfy the positive conditions 
\beq 
{\vec{\alpha}_{\cG_{1\times n}}}\cdot{\cM_{\cG_{n\times n}}}={\vec{h}_{\cG_{1\times n}}}\geqslant 0_{1\times n}, ~\forall \Delta_{\pi_i,\ell}\geqslant\Delta_{\pi_i,\ell}^*,~~ \label{Gpstv}
\eeq 
where $\Delta_{\pi_i,\ell}^*$  is the lowest allowed scaling dimension of spin $\ell$ operators in the $\pi_i$ representation given by either the unitary bound or certain assumption on the CFT spectrum. In particular the $SO(N)$ vector bootstrap linear functionals satisfy the positive conditions
\bea 
\vec{\alpha}_{SO(N)_{1\times 3}}\cdot\cM_{SO(N)_{3\times 3}}&=&(h_S, h_T, h_A)
\geqslant 0_{1\times 3},  \\
&&\forall~ \Delta_{S/T/A,\ell}\geqslant \Delta_{S/T/A,\ell}^*.~~~~~\label{SONpstv}
\eea 
Due to the algebraic relation (\ref{G2SO}), one can construct the linear functions $\vec{\alpha}_{\cG_{1\times n}}$ in (\ref{Gpstv}) from the $SO(N)$ linear functionals $\vec{\alpha}_{SO(N)_{1\times 3}}$:
\beq 
\vec{\alpha}_{\cG_{1\times n}}=\vec{\alpha}_{SO(N)_{1\times 3}}\cdot \mT_{\cG_{3\times n}} \label{sonfcl}
\eeq 
whose actions on the crossing equations $\cM_\cG$ are
\bea 
{\vec{\alpha}_{\cG_{1\times n}}}\cdot{\cM_{\cG_{n\times n}}}&=&(\vec{\alpha}_{SO(N)_{1\times 3}}\cdot \mT_{\cG_{3\times n}})\cdot{\cM_{\cG_{n\times n}}} \nn \\
&&\hspace{-2.7cm}=\vec{\alpha}_{SO(N)_{1\times 3}}\cdot \cM'_{SO(N)_{3\times n}} \cdot \textrm{diag}(1,x_{i},...y_{j},...)_{n\times n} ~~~~~~ \label{mixedaction}\\
&&\hspace{-2.7cm}=(h_S, x_{i}h_T,... y_j h_A,...)\geqslant 0_{1\times n}, \nn \\
&&\hspace{-1.2cm}\forall ~ \Delta_{S_\cG/R_i/r_j,\ell}
\geqslant
\Delta_{S_\cG/R_i/r_j,\ell}^*=\Delta_{S/T/A,\ell}^*, \label{Ggaps}
\eea
where $\cM'_{SO(N)_{3\times n}}$ is the $3\times n$ matrix in (\ref{G2SO}) related to $\cM_{SO(N)_{3\times 3}}$ with replicated  columns of $T,A$ sectors. The positivity of the recombination coefficients is critical for above actions being positive.

To summarize, due to the algebraic relation (\ref{G2SO}) with positive recombination coefficients, the $\cG$-symmetric crossing equations actually have the same positive algebraic structures of the $SO(N)$ vector's. Bounds from the $\cG$-symmetric bootstrap equations are saturated by the $SO(N)$-symmetric solutions unless the $SO(N)$ symmetric ``boundary conditions" in (\ref{Ggaps}) are violated explicitly.\footnote{The Eq. \ref{mixedaction} proves any spectrum excluded by the $SO(N)$ vector bootstrap is also excluded by the $\cG$-symmetric bootstrap, on the other hand, the $SO(N)$-symmetric solution can be directly decomposed into the $\cG$-symmetric solution. Therefore the two bootstrap setups should have the same excluded and allowed parameter spaces, i.e., they are identical.} 

\section{Examples of crossing equations with different global symmetries }\label{app:moregps}

CFTs with various types of global symmetries have been studied using modern conformal bootstrap approach \cite{Rattazzi:2008pe}. A comprehensive review on these studies can be found in \cite{Poland:2018epd}. The bootstrap crossing equations computed in these studies provide abundant examples to verify the proposed algebraic relation  between crossing equations with different global symmetries and $SO(N)$-ization. We have found the relation (\ref{G2SO}) and the positivity of recombination coefficients $x_i,y_j$ are supported by all these examples, providing compelling evidence for the main proposal (\ref{G2SO}-\ref{gnewGA}) of this work. Some of the physically interesting examples are presented in this section, while the algebraic relation of other theories and groups can be obtained similarly.
More examples can be found in an attached {\it Mathematica} file.

\subsection{Crossing equations of $SU(N)$ adjoint four-point correlators} \label{ceadj}
We show the relation between the crossing equations of $SU(N)$ adjoint  ($\cM_{SU(N)_\textrm{Adj}}$) and $SO(N^2-1)$ vector scalars.
Consider an $SU(N)$ adjoint scalar $\cO$, its OPE $\cO\times \cO$ contains 6 different sectors  
\beq 
\textrm{Adj}\otimes\textrm{Adj}\rightarrow S^+ \oplus\textrm{Adj}^+ \oplus\textrm{Adj}^- \oplus (A\bar{S}+S\bar{A})^-\oplus A\bar{A}^+\oplus S\bar{S}^+,
\eeq  
where $S, \textrm{Adj}$ denotes $SU(N)$ singlet and adjoint representations, while $A/S ~(\bar{A}/\bar{S})$ denote representations with anti-symmetric/symmetric of the $SU(N)$ fundamental (anti-fundamental) indices. The superscripts in $SU(N)$ representations correspond to even/odd spin selection rules.
We follow the convention in \cite{Berkooz:2014yda}.
Crossing equations $\cM_{SU(N)_\textrm{Adj}}$ can be written in following matrix form
\bea
&&\cM_{SU(N)_\textrm{Adj}}= \label{app:SUNceq}\\
&&\left(
\begin{array}{cccccc}
 0 & 0 & 0 & -F & F & F \\
 0 & \frac{2 F}{N} & 0 & 0 & -\frac{F}{N-2} & \frac{F}{N+2} \\
 0 & -F & -F & \frac{F}{N} & \frac{F}{N-2} & \frac{F}{N+2} \\
 F & -\frac{16 F}{N} & 0 & 0 & \frac{ 2 N^2F}{(N-1) (N-2)} & \frac{2 N^2 F}{(N+1) (N+2)} \\
 H & -\frac{4 H}{N} & 0 & -H & -\frac{N (N-3)H}{(N-1) (N-2)} & -\frac{N (N+3)H}{(N+1) (N+2)} \\
 0 & H & -H & \frac{H}{N} & \frac{(N-3)H}{N-2} & -\frac{ (N+3)H}{N+2} \\
\end{array}
\right), \nn
\eea
in which the columns are in the order
\beq
(S^+, ~\textrm{Adj}^+,~ \textrm{Adj}^-, ~(A\bar{S}+S\bar{A})^-,~A\bar{A}^+,~S\bar{S}^+), \label{app:vecs}
\eeq
In (\ref{app:SUNceq}) and below, we will omit the sub- and superscripts of correlation functions $F/H$ for simplicity. The corresponding symmetry representations and parity charges under the interchange $x_1\leftrightarrow x_2$ of each channel (columns in the matrix) are shown in a vector below the crossing equation, e.g. (\ref{app:vecs}).

The transformation matrix $\mT_{SU(N)_\textrm{Adj}}$ can be solved by imposing the $SO(N^2-1)\rightarrow SU(N)$ branching rules  and another constraint that there is no mixing between the functions $F$ and $H$ when mapping the $SU(N)$ adjoint crossing equations $\cM_{SU(N)_\textrm{Adj}}$ to the $SO(N^2-1)$ vector crossing equations $\cM_{SO(N^2-1)}$. The later constraint is justified as the functions $F$ and $H$ have opposite symmetries under  $u\leftrightarrow v$. These constraints lead to over-determined equations, which can be solved uniquely up to normalization
\bea
&&\mT_{SU(N)_\textrm{Adj}}= \\
&&\left(
\begin{array}{cccccc}
 1 & \frac{2 \left(N^4-2 N^2+2\right)}{N^4-N^2-2} & \frac{2 N}{N^2-2} & 0 & 0 & 0 \\
 -1 & \frac{-8 N^4+16 N^2+4}{-N^4+N^2+2} & -\frac{2 N}{N^2-2} & 1 & 0 & 0 \\
 0 & 0 & 0 & 0 & 1 & \frac{2 N}{N^2-2} \\
\end{array}
\right),  \nn
\eea
and the recombination coefficients are given by
{\small
\bea
 x_{\textrm{Adj}^+}&=&\frac{2 \left(N^4-5 N^2+4\right)}{N \left(N^4-N^2-2\right)}, x_{A\bar{A}^+}=\frac{(N-3) (N+1) N^2}{\left(N^2-2\right) \left(N^2+1\right)},  \nn\\
x_{S\bar{S}^+}&=&\frac{(N-1) (N+3) N^2}{\left(N^2-2\right) \left(N^2+1\right)},
~y_{\textrm{Adj}^-}=\frac{2 N}{N^2-2}, \label{app:xyinSUadj}\\ ~y_{A\bar{S}^-} &=&\frac{N^2-4}{N^2-2}.\nn
\eea
}

\subsection{Crossing equations of cubic model} \label{app:cubic}
The cubic model has discrete symmetry $C_N=\mathbb{Z}_2^N\rtimes S_N$ with scalars in the fundamental representation. 
Symmetry $C_N$ has close relation with $O(N)$. In particular the singlet and anti-symmetric representations of $O(N)$ construct the same irreducible representations of $C_N$, while $O(N)$ traceless symmetric representation decomposes into two irreducible representations of $C_N$, denoted by $V$ and $Y$ in \cite{Stergiou:2018gjj}.
Such $SO(N)\rightarrow C_N$ relation suggests the $SO(N)$-ized four-point correlator has the same spectrum as the cubic model in $S$ and $A$ sectors. Since the cubic model only has discrete global symmetry without a  spin 1 conserved current. Consequently there is no conserved current in the $SO(N)$-ized four-point correlator neither, though it is $SO(N)$ symmetric.
Following the convention in \cite{Stergiou:2018gjj}, 
the crossing equations of a fundamental scalar can be written in a matrix form
\bea
\cM_{C_N}=\left(
\begin{array}{cccc}
 0 & 0 & F & -F \\
 F & -\frac{2 F}{N} & F & F \\
 H & -\frac{2 H}{N} & -H & -H \\
 F & \left(2-\frac{2}{N}\right)F & 0 & 0 \\
\end{array}
\right),
\eea
Columns in above matrix are  in the order
\beq
(S^+, ~ V^+, ~ Y^+, ~ A^-).
\eeq
The  matrix $\mT_{C_N}$ which transforms $\cM_{C_N}$ to $\cM_{SO(N)}$ is
\bea
\mT_{C_N}=\left(
\begin{array}{cccc}
 \frac{N+1}{N+2} & -\frac{1}{N+2} & 0 & \frac{1}{N+2} \\
 -\frac{2}{N+2} & \frac{N}{N+2} & 0 & \frac{2}{N+2} \\
 0 & 0 & 1 & 0 \\
\end{array}
\right),
\eea
and the recombination coefficients are
\beq
x_{V^+}=\frac{2}{2+N},~ x_{Y^+}=\frac{N}{2+N}, ~ y_{A^-}=1.
\eeq

\subsection{Crossing equations of $SU(N)\times SU(N)$ bifundamental four-point correlators} \label{app:SU(N)-bf}
Many interesting 4D gauge theories have flavor symmetries $SU(N)\times SU(N)$, so this group could play an important role in our 4D bootstrap studies. The fermion bilinears $\cO_i^a\equiv\psi_{Li}\tilde{\psi}_R^a$ furnish bifundamental representation of the flavor symmetry. The $SU(N)\times SU(N)$ representations appearing in the OPE $\cO\times \cO$ are a direct product of the two $SU(N)$ representations $\{S, \textrm{Adj}, T, A\}$:
\bea 
\cO\otimes \cO\rightarrow
&& SS^+\oplus SS^-\oplus \textrm{AdjAdj}^+\oplus \textrm{AdjAdj}^-\oplus TA^- \nn\\ 
&&\oplus AA^+\oplus TT^+\oplus\textrm{Adj}S^+\oplus\textrm{Adj}S^-.
\eea 
There is a permutation symmetry between the two $SU(N)$ groups so we will not distinguish representations $LR$ and $RL$, e.g. $TA$ and $AT$.  
The crossing equations of bifundamental scalars are provided in \cite{Nakayama:2016knq}. Here we write the crossing equations into a compact matrix form
\begin{widetext}
\bea 
\cM_{SU(N)\times SU(N)}=\left(
\begin{array}{ccccccccc}
 F & F & \left(\frac{1}{N^2}+1\right)F &  \left(\frac{1}{N^2}+1\right)F & 0 & 0 & 0 & -\frac{2 F}{N} & -\frac{2 F}{N} \\
 H & H &  \left(\frac{1}{N^2}-1\right)H & \left(\frac{1}{N^2}-1\right) H& 0 & 0 & 0 & -\frac{2 H}{N} & -\frac{2 H}{N} \\
 0 & 0 & -\frac{2 F}{N} & -\frac{2 F}{N} & 0 & 0 & 0 & 2 F & 2 F \\
 0 & 0 & F & -F & -2F & F & F & 0 & 0 \\
 0 & 0 & -H & H & -2H & H & H & 0 & 0 \\
 F & -F & \frac{F}{N^2} & -\frac{F}{N^2} & 2F & F & F & -\frac{2 F}{N} & \frac{2 F}{N} \\
 H & -H & \frac{H}{N^2} & -\frac{H}{N^2} & -2H & -H & -H & -\frac{2 H}{N} & \frac{2 H}{N} \\
 0 & 0 & -\frac{F}{N} & \frac{F}{N} & 0 & -F & F & F & -F \\
 0 & 0 & -\frac{H}{N} & \frac{H}{N} & 0 & H & -H & H & -H \\
\end{array}
\right),
\eea
in which the columns are in the order
\beq
(SS^+,~SS^-, ~ \textrm{AdjAdj}^+,~ \textrm{AdjAdj}^-, ~TA^-,~ AA^+, ~TT^+, ~\textrm{Adj}S^+,~\textrm{Adj}S^- ),
\eeq
\end{widetext}
where $XY$ denotes representation $X(Y)$ for the left (right) $SU(N)$.
The crossing equations $\cM_{SU(N)\times SU(N)}$ can be transformed into $SO(2N^2)$ crossing equations by a linear transformation $\mT_{SU(N)\times SU(N)}$:
\begin{widetext}
\bea 
\mT_{SU(N)\times SU(N)} = 
\left(
\begin{array}{ccccccccc}
 \frac{2}{1-2 N^2} & 0 & -\frac{4 N}{2 N^4+N^2-1} & 2 & 0 & \frac{2}{2 N^2-1} & 0 & \frac{8 N^3}{2 N^4+N^2-1} & 0 \\
 \frac{2}{2 N^2-1}+2 & 0 & \frac{8 N^3}{2 N^4+N^2-1} & 0 & 0 & \frac{2}{1-2 N^2}+2 & 0 & \frac{8 N \left(N^2-1\right)}{2 N^4+N^2-1} & 0 \\
 0 & \frac{2}{1-2 N^2}+2 & 0 & 0 & \frac{4}{1-2 N^2} & 0 & \frac{2}{2 N^2-1}+2 & 0 & \frac{8 N}{2 N^2-1} \\
\end{array}
\right), \nn
\eea
and the recombination coefficients are
\bea
 x_{\textrm{AdjAdj}^+}&=&\frac{\left(N^2-1\right)^2}{2 N^4+N^2-1}, ~x_{TT^+}=\frac{N^2 (N+1)^2}{2 N^4+N^2-1}, ~x_{\textrm{Adj}S^+}=\frac{2 N \left(N^2-1\right)}{2 N^4+N^2-1}, 
 x_{ AA^+}=\frac{(N-1)^2 N^2}{2 N^4+N^2-1}, \\ ~y_{SS^-}&=&\frac{1}{2 N^2-1}, ~~~~~y_{ \textrm{AdjAdj}^-}=\frac{\left(N^2-1\right)^2}{N^2 \left(2 N^2-1\right)}, ~~~~
y_{TA^-}=\frac{2(N^2-1)}{2 N^2-1}, ~~~\, y_{\textrm{Adj}S^-}=\frac{2-2 N^2}{N-2 N^3}. 
\eea
\end{widetext}

\subsection{Crossing equations of $SO(N)\times SO(M)$ bifundamental four-point correlators} \label{app:SO(N)SO(M)-bf}
CFTs with $SO(N)\times SO(M)$ symmetry can be used to describe the phase transitions in frustrated spin models \cite{1998JPCM10, Delamotte:2003dw}, in which the order parameters construct bi-fundamental representations of the symmetry. Conformal bootstrap study of this theory has been initiated in \cite{Nakayama:2014lva}. 
The OPE of an $SO(N)\times SO(M)$ bifundamental scalar $\cO_{i,j}$ is given by:
\bea 
\cO\times\cO\rightarrow &&
SS^+\oplus ST^+\oplus  SA^-\oplus  TS^+ \\ &&\oplus TT^+\oplus  TA^-\oplus AS^-\oplus AT^-\oplus AA^+, \nn
\eea 
where the $S,T,A$ are the singlet, traceless symmetric and anti-symmetric representations of $SO(N)$ and $SO(M)$, and in the representation $XY$,  $X(Y)$ denotes the representation for the $SO(N)$ ($SO(M)$). 
The crossing equations of $SO(N)\times SO(M)$ bifundamental scalars are provided in \cite{Caracciolo:2014cxa}, which can be written in a compact matrix form
\begin{widetext}
\bea
\cM_{SO(N)\times SO(M)}=\left(
\begin{array}{ccccccccc}
 F & -\frac{2 F}{M} & 0 & -\frac{2 F}{N} &  \left(\frac{4}{N M}+1\right)F & F & 0 & F & F \\
 H & -\frac{2 H}{M} & 0 & -\frac{2 H}{N} & \left(\frac{4}{N M}-1\right)H & -H & 0 & -H & -H \\
 0 & 0 & 0 & F &  \left(1-\frac{2}{M}\right)F & F & -F & \left(\frac{2}{M}-1\right)F & -F \\
 0 & 0 & 0 & -H &  \left(\frac{2}{M}+1\right)H & H & H & \left(-\left(\frac{2}{M}+1\right)\right)H & -H \\
 0 & F & -F & 0 &  \left(1-\frac{2}{N}\right)F &  \left(-\left(1-\frac{2}{N}\right)\right)F & 0 & F & -F \\
 0 & -H & H & 0 &  \left(\frac{2}{N}+1\right)H &  \left(-\left(\frac{2}{N}+1\right)\right)H & 0 & H & -H \\
 0 & -F & -F & -F & \left(\frac{2}{M}+\frac{2}{N}\right)F & \frac{2 F}{N} & -F & \frac{2 F}{M} & 0 \\
 0 & H & H & -H &  \left(\frac{2}{M}-\frac{2}{N}\right)H & -\frac{2 H}{N} & -H & \frac{2 H}{M} & 0 \\
 0 & 0 & 0 & 0 & F & -F & 0 & -F & F \\
\end{array}
\right),
\eea
in which the columns are in the order
\beq
(SS^+,~ST^+, ~ SA^-,~ TS^+, ~TT^+,~ TA^-, ~AS^-, ~AT^-,~AA^+ ),
\eeq
\end{widetext}
Note with $N=M$ the two $SO(N)$'s are interchangeable and the representations like $TA$ and $AT$ are indistinguishable, which can result in different crossing equations.
The crossing equations $\cM_{SO(N)\times SO(M)}$ can be transformed into $SO(MN)$ crossing equations by a linear transformation $\mT_{SO(N)\times SO(M)}$:
\begin{widetext}
\bea 
\mT_{SO(N)\times SO(M)} =
\left(
\begin{array}{ccccccccc}
 0 & 0 & \frac{(N-1) (M N+M+1)}{(M N-1) (M N+2)} & 0 & \frac{(M-1) (M N+N+1)}{(M N-1) (M N+2)} & 0 & -\frac{(M-1) (N-1)}{(M N-1) (M N+2)} & 0 & 1 \\
 1 & 0 & \frac{(M-1) (M+2) N}{(M N-1) (M N+2)} & 0 & \frac{M (N-1) (N+2)}{(M N-1) (M N+2)} & 0 & -\frac{M^2 N+M N^2-2}{(M N-1) (M N+2)} & 0 & 0 \\
 0 & 1 & 0 & -\frac{M-1}{M N-1} & 0 & -\frac{N-1}{M N-1} & 0 & -\frac{M-N}{M N-1} & 0 \\
\end{array}
\right), 
\eea
and the recombination coefficients are
{\small
\bea
x_{ST^+} &= \frac{\left(M^2+M-2\right) N}{M^2 N^2+M N-2},x_{TS^+} &= \frac{M \left(N^2+N-2\right)}{M^2 N^2+M N-2},
~~~~~~x_{TT^+}= \frac{\left(M^2+M-2\right) \left(N^2+N-2\right)}{M^2 N^2+M N-2}, y_{SA^-}= \frac{M-1}{M N-1}, \\
x_{AA^+} &= \frac{(M-1) M (N-1) N}{M^2 N^2+M N-2}, 
y_{TA^-} &= \frac{(M-1) \left(N^2+N-2\right)}{N (M N-1)},y_{AT^-}= \frac{\left(M^2+M-2\right) (N-1)}{M (M N-1)}, ~~~~~~~~y_{AS^-}= \frac{N-1}{M N-1}.~~
\eea }
\end{widetext}

\section{$SO(N)$-ization of generalized free theories} \label{App-GFTs}

In this section we provide two examples for the $SO(N)$-ization: the four-point correlators of the $SU(N)$ fundamental generalized free scalars and the $SU(N)$ adjoint generalized free fermion bilinears. 

Let us consider an $SU(N)$ fundamental scalar  $\Phi^i$, whose four-point correlator can be solved in generalized free field theory  {\small
\beq
\langle \Phi^i(x_1)\Phi^\dagger_{\bar{i}}(x_2)\Phi^j(x_3)\Phi^\dagger_{\bar{j}}(x_4)\rangle=\frac{1}{x_{12}^{2\Delta_\Phi}x_{34}^{2\Delta_\Phi}}
(\delta^i_{\bar{i}}\delta^j_{\bar{j}}+\delta^i_{\bar{j}}\delta^j_{\bar{i}}u^{\Delta_\Phi}). \label{SU1}
\eeq }
There are 6 invariant tensors in the $SU(N)$ crossing equations (\ref{crqSU}) and the associated correlation functions $G_i(u,v)$  can be solved through the four-point correlation function (\ref{SU1}) and its modified configurations $\langle \Phi^i\Phi^\dagger_{\bar{i}}\Phi^\dagger_{\bar{j}}\Phi^j\rangle$, $\langle \Phi^i\Phi^j\Phi^\dagger_{\bar{i}}\Phi^\dagger_{\bar{j}}\rangle$: {\small
\bea
G_S^+ &=&1+\frac{1}{2N}u^{\Delta_\phi}\left(v^{-\Delta_\phi}+1\right),  \\
G_ {\text{Adj}}^+ &=& G_T^+=\frac{1}{2}u^{\Delta_\phi}\left(v^{-\Delta_\phi}+1\right), \\
G_{\text{Adj}}^- &=& G_A^-=N G_S^-=\frac{1}{2}u^{\Delta_\phi} \left(v^{-\Delta_\phi}-1\right).  \label{GFTSU}
\eea}
Applying the above solutions in (\ref{newGS}-\ref{newGA}) we can reproduce the exact $SO(2N)$ correlation functions (\ref{cAphi})
\bea
G_S^+ &=&1+\frac{1}{N}u^{\Delta_\phi}\left(1+v^{-\Delta_\phi}\right),  \\
G_T^+ &=& \frac{1}{2}u^{\Delta_\phi}\left(1+v^{-\Delta_\phi}\right),  \\
G_A^- &=&\frac{1}{2}u^{\Delta_\phi} \left(-1+v^{-\Delta_\phi}\right), \label{GFTON}
\eea  
which can be obtained from Wick contractions of the $SO(N)$ vector four-point correlator.
This is consistent but trivial -- the $SO(2N)$ symmetry enhancement of the $SU(N)$ fundamental generalized free field theory is obvious.

A less trivial example is provided by the four-point correlator of the 3D fermion bilinear operator $\cO_i^m\equiv\bar{\psi}_i\psi^m-\frac{1}{N}\delta_i^m\bar{\psi}_k\psi^k$, which has scaling dimension $2\Delta$ and furnishes the adjoint representation of the $SU(N)$ flavor symmetry. This is a real representation of $SU(N)$, therefore it suffices to consider the single correlator $\langle \cO_i^m\cO_j^n\cO_k^p\cO_l^q\rangle$ only.
The crossing equations of this correlator have been studied using conformal bootstrap \cite{Berkooz:2014yda, Iha:2016ppj, Li:2018lyb}.  For $N\geqslant 4$ there are 6  sectors in the crossing equations $\cM_{SU(N)_\textrm{Adj}}$, which are shown in Section  \ref{ceadj} and  explicit formulas of the $SU(N)$-invariant four-point tensors $\cT_{\pi_i}$ are provided in \cite{Berkooz:2014yda, Iha:2016ppj}. 
The fermion bilinear four-point correlator is 
\begin{widetext}
 \bea
\mG_{SU(N)-\textrm{adj}} = \left( \cT_S^+ G_S^++ \cT_{\textrm{Adj}}^+G_{\textrm{Adj}}^++   \cT_{\textrm{Adj}}^-G_{\textrm{Adj}}^-  +(\cT_{A\bar{S}}+\cT_{S\bar{A}})^-G_{A\bar{S}}^-+
\cT_{A\bar{A}}^+G_{A\bar{A}}^++\cT_{S\bar{S}}^+G_{S\bar{S}}^+\right).
\eea 
In the generalized free fermion theory, contributions to the correlator from disconnected diagrams are  
\bea 
G_S^+ &=1+\frac{u^{2 \Delta } \left(v^{-2 \Delta }+1\right)}{N^2-1},  G_{\textrm{Adj}}^+ &=\frac{N u^{2 \Delta } \left(v^{-2 \Delta }+1\right)}{2 \left(N^2-4\right)}, 
G_{\textrm{Adj}}^- =\frac{u^{2 \Delta } \left(v^{-2 \Delta }-1\right)}{2 N},   \\
G_{A\bar{S}}^- &=\frac{1}{4} u^{2 \Delta } \left(v^{-2 \Delta }-1\right),  
G_{A\bar{A}}^+ &= \frac{1}{4} u^{2 \Delta } \left(v^{-2 \Delta }+1\right), 
~\,G_{S\bar{S}}^+ =\frac{1}{4} u^{2 \Delta } \left(v^{-2 \Delta }+1\right), \label{disc}
\eea
and the contributions from connected diagrams are
\bea
G_S^+ &=&\frac{2u^{\Delta-\frac{1}{2}} v^{-\Delta-\frac{1}{2}} }{N \left(N^2-1\right)  } \left(N^2 \left((v-1) \left(v^{\Delta +\frac{1}{2}}-1\right)-u \left(v^{\Delta +\frac{1}{2}}+1\right)\right) \right.  \label{connect1} \nn\\
&& \hspace{2.5cm} \left. -u^{\Delta +\frac{3}{2}}+(v+1) u^{\Delta +\frac{1}{2}}+u v^{\Delta +\frac{1}{2}}+u-(v-1) v^{\Delta +\frac{1}{2}}+v-1\right),  \\
G_{\textrm{Adj}}^+&=&-\frac{u^{\Delta-\frac{1}{2} } v^{-\Delta -\frac{1}{2}}}{2 \left(N^2-4\right) }\left(\left(N^2-4\right) \left(u \left(v^{\Delta +\frac{1}{2}}+1\right)-  (v-1) \left(v^{\Delta +\frac{1}{2}}-1\right)\right)+4 u^{\Delta +\frac{3}{2}}-4 (v+1) u^{\Delta +\frac{1}{2}}\right), \\
G_{\textrm{Adj}}^-&=&\frac{1}{2} u^{\Delta -\frac{1}{2}} v^{-\Delta -\frac{1}{2}} \left(u \left(v^{\Delta +\frac{1}{2}}-1\right)-(v-1) \left(v^{\Delta +\frac{1}{2}}+1\right)\right), \\
G_{A\bar{A}}^+ &=&-G_{S\bar{S}}^+=\frac{1}{2} u^{2 \Delta } v^{-\Delta -\frac{1}{2}} (-u+v+1), ~~~~G_{A\bar{S}}^-=0. \label{connect2}
\eea
\end{widetext}
The recombination coefficients solved from the crossing equations $\cM_{SU(N)_\textrm{Adj}}$ are provided in (\ref{app:xyinSUadj}).
Using the formulas (\ref{gnewGS}-\ref{gnewGA}), we can construct $SO(N^2-1)$-symmetric correlation functions straightforwardly. In particular the disconnected  correlation functions (\ref{disc}) reproduce the $SO(N^2-1)$ correlator of generalized free field theory (\ref{GFTON}). The connected correlation functions (\ref{connect1}-\ref{connect2}) lead to more complicated results
\begin{widetext}
\bea
G'_S&=&\frac{u^{\Delta -\frac{1}{2}} v^{-\Delta -\frac{1}{2}} }{N \left(N^2-2\right) \left(N^2+1\right)}\left(N^4 \left(2 u^{\Delta +\frac{3}{2}}-2 (v+1) u^{\Delta +\frac{1}{2}}-u \left(v^{\Delta +\frac{1}{2}}+1\right)+(v-1) \left(v^{\Delta +\frac{1}{2}}-1\right)\right)\right.  \nn\\
&&\hspace{3.3cm}\left.+4 \left(u^{\Delta +\frac{3}{2}}-(v+1) u^{\Delta +\frac{1}{2}}-u \left(v^{\Delta +\frac{1}{2}}+1\right)+(v-1) \left(v^{\Delta +\frac{1}{2}}-1\right)\right) \right. 
\nn\\
&&\hspace{3.3cm}\left.+N^2 \left(-4 u^{\Delta +\frac{3}{2}}+4 (v+1) u^{\Delta +\frac{1}{2}}+5 u \left(v^{\Delta +\frac{1}{2}}+1\right)-5 (v-1) \left(v^{\Delta +\frac{1}{2}}-1\right)\right)\right), \label{app:fbS}\\
G'_T&=&\frac{2 u^{\Delta -\frac{1}{2}} v^{-\Delta -\frac{1}{2}}}{N \left(N^2-1\right)} \left(-u^{\Delta +\frac{3}{2}}+(v+1) u^{\Delta +\frac{1}{2}}+u v^{\Delta +\frac{1}{2}}+u-(v-1) v^{\Delta +\frac{1}{2}}+v-1\right. \nn\\
&& \hspace{2.6cm}\left.+N^2 \left((v-1) \left(v^{\Delta +\frac{1}{2}}-1\right)-u \left(v^{\Delta +\frac{1}{2}}+1\right)\right)\right), \label{app:fbT}\\
G'_A&=&\frac{N u^{\Delta -\frac{1}{2}} v^{-\Delta -\frac{1}{2}} \left(u \left(v^{\Delta +\frac{1}{2}}-1\right)-(v-1) \left(v^{\Delta +\frac{1}{2}}+1\right)\right)}{N^2-2}.\label{app:fbA}
\eea
\end{widetext}
The above correlation functions  do satisfy the $SO(N^2-1)$ crossing equations (\ref{crqON})!

\section{$SO(N)$-ization of WZW model} \label{app:WZW}
A more non-trivial test of the $SO(N)$-ization mechanism is provided by the 2D Wess-Zumino-Witten (WZW) model \cite{Wess:1971yu, Witten:1983tw}.
Let us consider the four-point correlator of the fundamental field $g_{\alpha}^{\ad}$ in the $SU(N_f)_k$ WZW model: 
\bea  \label{wzw4pt}
\langle g_{\alpha}^{\ad}(z_1,\bar{z}_1)g^{\beta}_{\bd}(z_2,\bar{z}_2)^{-1}g_{\rho}^{\rd}(z_3,\bar{z}_3)g_{\sid}^{\sigma}(z_4,\bar{z}_4)^{-1}\rangle=& \nn\\
& \hspace{-6cm}|z_{12}|^{-2\Delta_g}|z_{34}|^{-2\Delta_g}  \mG_{\alpha \bd\rho \sid}^{\ad\beta \rd\sigma}(\eta,\bar{\eta}),
\eea
where $z=x^0+i x^1$, $z_{ij}=z_i-z_j$, $\eta=z_{12}z_{34}/z_{13}z_{24}$ and $\bar{z}, \bar{\eta}$ are their complex conjugates. The conformal primary field $g_{\alpha}^{\ad}$ forms bi-fundamental representation of the symmetry $SU(N_f)_L\times SU(N_f)_R$.
The correlation function $\mG$ has been solved in \cite{Knizhnik:1984nr} using the  Knizhnik–Zamolodchikov equation and crossing symmetry \footnote{The crossing equations of 2D WZW models have a subtle difference comparing with higher dimensional crossing equations. In 2D WZW model, some of the components in the four-point correlator (\ref{wzw4pt}), such as $G_{T,A}^-$, $G_{\textrm{Adj},S}^\pm $ are not symmetric between variables $\eta$ and $\bar{\eta}$. In contrast, the higher dimensional conformal blocks are symmetric between the two variables \cite{Dolan:2000ut,Dolan:2003hv}.}
\beq \label{wzwcr}
|1-\eta|^{2\Delta_g} \mG_{\alpha \bd\rho \sid}^{\ad\beta \rd\sigma}(\eta,\bar{\eta})=
|\eta|^{2\Delta_g}
\mG_{\alpha \sid\rho \bd}^{\ad\sigma \rd\beta}(1-\eta,1-\bar{\eta}),
\eeq 
which is:
\begin{widetext}
\bea
\mG_{\alpha \bd\rho \sid}^{\ad\beta \rd\sigma}(\eta,\bar{\eta})= \left(\delta_\alpha^\beta\delta_\rho^\sigma P_1(\eta)+\delta_\alpha^\sigma\delta_\rho^\beta P_2(\eta) \right)
\left(\delta^{\ad}_{\bd}\delta^{\rd}_{\sid} P_1(\bar{\eta})+\delta^{\rd}_{\bd}\delta^{\ad}_{\sid} P_2(\bar{\eta}) \right)+ h\left(\delta_\alpha^\beta\delta_\rho^\sigma Q_1(\eta)+\delta_\alpha^\sigma\delta_\rho^\beta Q_2(\eta) \right)
\left(\delta^{\ad}_{\bd}\delta^{\rd}_{\sid} Q_1(\bar{\eta})+\delta^{\rd}_{\bd}\delta^{\ad}_{\sid} Q_2(\bar{\eta}) \right), \nn\\ \label{wzw:solution}
\eea
and the functions $P_i, Q_i$ are given by
\bea
P_1(\eta)&=&(1-\eta)^{h_A-2h_g} \,_2F_1(\frac{1}{N_f+k},-\frac{1}{N_f+k},1-\frac{N_f}{N_f+k},\eta), \\
P_2(\eta)&=&\frac{1}{k}\eta(1-\eta)^{h_A-2h_g} \,_2F_1(1+\frac{1}{N_f+k},1-\frac{1}{N_f+k},2-\frac{N_f}{N_f+k},\eta), \\
Q_1(\eta)&=&\eta^{h_A}(1-\eta)^{h_A-2h_g} \,_2F_1(\frac{N_f-1}{N_f+k},\frac{N_f+1}{N_f+k},1+\frac{N_f}{N_f+k},\eta), \\
Q_2(\eta)&=&-N_f\,\eta^{h_A}(1-\eta)^{h_A-2h_g} \,_2F_1(\frac{N_f-1}{N_f+k},\frac{N_f+1}{N_f+k},\frac{N_f}{N_f+k},\eta). 
\eea 
\end{widetext}
The scaling dimensions  $h_g, h_A$ and the coefficient $h$ are 
\bea 
h_g &=&\frac{N_f^2-1}{2N_f(N_f+k)},~~
h_A=\frac{N_f}{N_f+k}, \\
h&=&\frac{1}{N_f^2}\frac{\Gamma[\frac{N_f-1}{N_f+k}]\Gamma[\frac{N_f+1}{N_f+k}]\Gamma[\frac{k}{N_f+k}]^2}{\Gamma[\frac{k+1}{N_f+k}]\Gamma[\frac{k-1}{N_f+k}]\Gamma[\frac{N_f}{N_f+k}]^2}.
\eea 

The four-point correlation function (\ref{wzw:solution}) consists of two  parts, in which the $SU(N_f)_L$ and $SU(N_f)_R$ indices are factorized.
It is straightforward to decompose the correlation function in terms of $SU(N_f)_L$ and $SU(N_f)_R$ representations ($\pi_L$ and $\pi_R$) $G_{\pi_L\pi_R}^\pm$ with definite parity charges under the interchange of external coordinates $x_1\leftrightarrow x_2$. Here $\pi_{L,R}$ are those appearing in $\cM_{SU(N)}$ (\ref{crqSU}).  Since the conformal primary operator $g$ constructs bi-fundamental representation of $SU(N_f)_L\times SU(N_f)_R$, its crossing equations are related to the crossing equations in (\ref{app:SU(N)-bf}). However, there is a subtle difference between the crossing equations of 2D WZW model and higher dimensional CFTs. In the later case the CPWs and conformal blocks are symmetric between variables $\eta, \bar{\eta}$, while several $SU(N_f)_L\times SU(N_f)_R$ representations in (\ref{wzw:solution}) are not, thought such symmetry is restored in the whole four-point correlator, associated with exchanging of $SU(N_f)_L$ and $SU(N_f)_R$ indices.
Due to an algebraic relation similar to (\ref{SU2SO}), the $SU(N_f)\times SU(N_f)$ bifundamental crossing equations  can be linearly transformed to the $SO(2N_f^2)$ vector crossing equations associated with the $SO(2N_f^2)\rightarrow SU(N_f)_L\times SU(N_f)_R$ branching rules 
\begin{widetext}
\bea
SO(2N_f^2) &~~~\longrightarrow ~~~&  SU(N_f)_L\times SU(N_f)_R \nn\\
S &\longrightarrow &(S,S), \\
T &\longrightarrow & (\textrm{Adj},\textrm{Adj})\oplus(\textrm{Adj},S)\oplus(S,\textrm{Adj})\oplus(T,T)\oplus(A,A),~~~~ \\
A &\longrightarrow & (S,S)\oplus(\textrm{Adj},\textrm{Adj})\oplus(\textrm{Adj},S)\oplus(S,\textrm{Adj})\oplus(T,A)\oplus(A,T),
\eea
\end{widetext}
and $SO(2N_f^2)$-symmetric correlation functions ${G'}_{S/T/A}$ can be constructed following (\ref{gnewGS}-\ref{gnewGA}) with recombination coefficients $x_i, y_j$ given in Section \ref{app:SU(N)-bf}.
 The correlation functions with general $N_f, k>1$ are rather complicated. For the $k=1$ case, which corresponds to the IR fixed point of  2D Quantum Electrodynamics (QED$_2$), the factor $h$ in (\ref{wzw:solution}) vanishes and the four-point function is simplified significantly. The $SO(2N_f^2)$-ization correlation function is
 \begin{widetext}
\bea 
G'_S(u,v)&=&\frac{v^{-\frac{1}{N_f}-1} }{2 N_f^2}\left(v^{2/N_f} (-N_f u+(N_f-1) N_f v+N_f+u)+v (N_f (N_f-u+v-1)+u)\right), \label{app:wzwS}\\
G'_T(u,v)&=&\frac{1}{4 \left(2 N_f^4+N_f^2-1\right) v}
\left(N_f^2 (N_f+1)^2 u^{2-\frac{2}{N_f}} v^{\frac{1}{N_f}}+2 \left(N_f^2-1\right)^2 u \left(v^{\frac{1}{N_f}}+v^{1-\frac{1}{N_f}}\right)-(N_f-1)^2 N_f^2 u^{1-\frac{2}{N_f}}  \right. \nn\\
&&\left. \times v^{\frac{1}{N_f}}(u-2 (v+1))+2 \left(N_f^2-1\right) \left(v^{\frac{1}{N_f}} (2 u-N_f (u+v-1))+v^{1-\frac{1}{N_f}} (N_f (-u+v-1)+2 u)\right)\right), \label{app:wzwT}\\
G'_A(u,v)&=&\frac{u^{-\frac{2}{N_f}} v^{-\frac{N_f+1}{N_f}}}{2 \left(2 N_f^2-1\right)}
\left(-v\, u^{\frac{2}{N_f}} \left(N_f \left(u \left(N_f-1\right)+v-1\right)+1\right)+v^{\frac{2}{N_f}} \left(-u (v-1) \left(N_f^2-1\right)+\right.\right. \nn\\
&& \hspace{2.3cm}\left.\left.u^{\frac{2}{N_f}} \left(-N_f (u+v-1)+u N_f^2+v\right)\right)\right).\label{app:wzwA}
\eea 
\end{widetext}
In the large $N_f$ limit, the above correlation functions contain leading $u,v$ analytical terms: 
\bea 
G'_S(u,v) &=& 1+\frac{-u v-u+v^2-2 v+1}{2 {N_f} v}, \label{freefermionS}\\
G'_T(u,v) &=& \frac{u+ u v}{2 v} +\frac{2 u^2-3 u v-3 u+(v-1)^2}{4N_f v},~~ \label{freefermionT}\\
G'_A(u,v) &=& \frac{u - u v}{2 v}+\frac{u v-u-v^2+1}{4N_f v},\label{freefermionA}
\eea 
consistent with the large $N$ limit of the $SO(N)$-ization of the 2D free fermion bilinear four-point correlators (\ref{app:fbS}-\ref{app:fbA}). This is expected as the conformal primary operator $g_{\alpha}^{\ad}$ in the $SU(N_f)_{k=1}$ WZW model is equivalent to the free fermion bilinears in the large $N_f$ limit. 
The four-point function (\ref{wzw4pt}) can also be truncated into subgroups of $SU(N_f)_L\times SU(N_f)_R$ to realize different $SO(N)$-ization. For instance, one may take the diagonal part of $SU(N_f)_L\times SU(N_f)_R$ in (\ref{wzw4pt}) and construct an $SO(2N_f)$-symmetric four-point correlator. 
Overall our result suggests there are rich $SO(N)$-symmetric four-point correlation functions from the $SO(N)$-ization of the WZW model.

An interesting question is whether certain $SO(N)$-ized four-point correlation functions can play roles in determining the $SO(N)$ vector bootstrap bounds. In \cite{Li:2018lyb} a new family of bootstrap kinks have been discovered  in the 3D $SO(N)$ vector bootstrap bounds besides the well-known kinks \cite{Kos:2013tga} corresponding to the classical Wilson-Fisher $O(N)$ fixed points. These kinks also appear in the $SO(N)$ vector bootstrap bounds in general dimensions \cite{Li:2020bnb} and they approach free fermion theory in the large $N$ limit. There are evidence indicating they may correspond to gauge theories coupled with fermions, such as the IR fixed points of QED$_3$ \cite{Appelquist:1988sr, Nash:1989xx} and the Caswell-Banks-Zaks fixed points of 4D Quantum Chromodynamics  \cite{Caswell:1974gg, Banks:1981nn}. While the conformal gauge theories are non-$SO(N)$ symmetric and it was unclear how such theories could appear in the $SO(N)$ vector bootstrap bounds. The $SO(N)$-ization provides a possible explanation for this puzzle. The $SU(N_f)_k$  WZW model can be considered as dimensional continuations of the IR fixed points of higher dimensional $U(k)$ gauge theories coupled to $N_f$ flavors of Dirac fermions \cite{Witten:1983ar, Gepner:1984au}. We hope the $SO(N)$-ization of WZW model could provide a key to decode the $SO(N)$ vector bootstrap results.

\section{Outlooks}
The algebraic property of the crossing equations studied in this work clarifies a fundamental barrier for the widely interested modern conformal bootstrap program, which aims to numerically solve or classify the CFTs using crossing equations and positivity condition. This ambitious project relies on the assumption that the dynamics of CFTs is encoded in the crossing equations and positivity condition.
However, our results show that for various types of symmetries $\cG$, the $\cG$-symmetric four-point crossing equations are actually equipped with an $SO(N)$ symmetric positive structure. According to the conformal bootstrap algorithm, without $SO(N)$ symmetry breaking assumptions on the CFT data, the $\cG$-symmetric conformal bootstrap  degenerates to the $SO(N)$ vector bootstrap, which provides strong restrictions to numerically solve the non-$SO(N)$ symmetric theories. The conclusion is that due to a dedicate algebraic relation of the conformal four-point crossing equations, the $\cG$-symmetric crossing equations and positivity condition alone do not contain specific information for the non-$SO(N)$ symmetric theories, and it needs to introduce new ingredients or specific assumptions on the spectrum to move further.\footnote{
These restrictions should not be considered as a  no-go theorem for the bootstrap program.  One may introduce extra assumptions on the CFT data of the target CFTs in the bootstrap implementations, which can break the $SO(N)$ symmetry explicitly. Effects of these assumptions vary for different theories and need to be studied case by case.}

We discuss two interesting implications of our results:
 
{\it{$6j$-symbols, crossing symmetry and positivity:}}  
Our results only uncover a tip of the iceberg for a deep connection between representation theory and CFTs. The conformal bootstrap approach exploits constraints from the conformal blocks, crossing symmetry and positivity, and its remarkable success suggests there are unknown positive structures in the conformal blocks associated with crossing symmetry, i.e. the $6j$ symbols of the conformal group $SO(D+1,1)$ \cite{Gadde:2017sjg, Liu:2018jhs}.  The symmetry properties of the conformal four-point functions provide reminiscent while much simpler examples on the positive structures in the $6j$ symbols. We have verified this symmetry property for a variety of groups, and it would be very instructive to obtain a mathematically general proof for the existence of these linear transformations and the positivity of the recombination coefficients. This can clarify how the positivity arises in the $6j$ symbols of the {\it compact global} symmetry groups, which can provide insights to study the positive structures in the $6j$ symbols of the {\it non-compact local} conformal group $SO(D+1,1)$.
 
{\it Constraints on the RG flows:}
In 3D the $SO(N)$ vector bootstrap bounds are  saturated by the critical $O(N)$ vector models \cite{Kos:2013tga}. Considering there are abundant $SO(N)$-symmetric four-point correlators (\ref{gnewGS}-\ref{gnewGA}) constructed   from the scalar CFTs with symmetries $\cG\subsetneq SO(N)$ \cite{Osborn:2017ucf, Rychkov:2018vya}, the bootstrap results indicate these  non-$SO(N)$ symmetric scalar CFTs contain certain factors which make their $SO(N)$-ized correlators locating in the bulk of the $SO(N)$ vector bootstrap bounds. This would further suggest the RG flows from the critical $O(N)$ vector models to non-$SO(N)$ symmetric fixed points driven by the $SO(N)$ symmetry breaking couplings are restricted within certain directions.

\section*{Acknowledgements}
It is a pleasure to thank David Poland for stimulating discussions and collaborations on relevant projects. 
The author is grateful to David Poland and Slava Rychkov for valuable comments and suggestions on the draft.
The author would like to thank Soner Albayrak and Rajeev Erramilli for discussions.
The work of ZL  is supported by Simons Foundation grant 488651 (Simons Collaboration on the Nonperturbative Bootstrap) and DOE grant no.\ DE-SC0020318. 

\bibliography{Symin4pt}

\providecommand{\noopsort}[1]{}\providecommand{\singleletter}[1]{#1}%
\begin{thebibliography}{44}%
\makeatletter
\providecommand \@ifxundefined [1]{%
 \@ifx{#1\undefined}
}%
\providecommand \@ifnum [1]{%
 \ifnum #1\expandafter \@firstoftwo
 \else \expandafter \@secondoftwo
 \fi
}%
\providecommand \@ifx [1]{%
 \ifx #1\expandafter \@firstoftwo
 \else \expandafter \@secondoftwo
 \fi
}%
\providecommand \natexlab [1]{#1}%
\providecommand \enquote  [1]{``#1''}%
\providecommand \bibnamefont  [1]{#1}%
\providecommand \bibfnamefont [1]{#1}%
\providecommand \citenamefont [1]{#1}%
\providecommand \href@noop [0]{\@secondoftwo}%
\providecommand \href [0]{\begingroup \@sanitize@url \@href}%
\providecommand \@href[1]{\@@startlink{#1}\@@href}%
\providecommand \@@href[1]{\endgroup#1\@@endlink}%
\providecommand \@sanitize@url [0]{\catcode `\\12\catcode `\$12\catcode
  `\&12\catcode `\#12\catcode `\^12\catcode `\_12\catcode `\%12\relax}%
\providecommand \@@startlink[1]{}%
\providecommand \@@endlink[0]{}%
\providecommand \url  [0]{\begingroup\@sanitize@url \@url }%
\providecommand \@url [1]{\endgroup\@href {#1}{\urlprefix }}%
\providecommand \urlprefix  [0]{URL }%
\providecommand \Eprint [0]{\href }%
\providecommand \doibase [0]{https://doi.org/}%
\providecommand \selectlanguage [0]{\@gobble}%
\providecommand \bibinfo  [0]{\@secondoftwo}%
\providecommand \bibfield  [0]{\@secondoftwo}%
\providecommand \translation [1]{[#1]}%
\providecommand \BibitemOpen [0]{}%
\providecommand \bibitemStop [0]{}%
\providecommand \bibitemNoStop [0]{.\EOS\space}%
\providecommand \EOS [0]{\spacefactor3000\relax}%
\providecommand \BibitemShut  [1]{\csname bibitem#1\endcsname}%
\let\auto@bib@innerbib\@empty
\bibitem [{\citenamefont {Polyakov}(1974)}]{Polyakov:1974gs}%
  \BibitemOpen
  \bibfield  {author} {\bibinfo {author} {\bibfnamefont {A.~M.}\ \bibnamefont
  {Polyakov}},\ }\bibfield  {title} {\bibinfo {title} {{Nonhamiltonian approach
  to conformal quantum field theory}},\ }\href@noop {} {\bibfield  {journal}
  {\bibinfo  {journal} {Zh. Eksp. Teor. Fiz.}\ }\textbf {\bibinfo {volume}
  {66}},\ \bibinfo {pages} {23} (\bibinfo {year} {1974})}\BibitemShut {NoStop}%
\bibitem [{\citenamefont {Ferrara}\ \emph {et~al.}(1973)\citenamefont
  {Ferrara}, \citenamefont {Grillo},\ and\ \citenamefont
  {Gatto}}]{Ferrara:1973yt}%
  \BibitemOpen
  \bibfield  {author} {\bibinfo {author} {\bibfnamefont {S.}~\bibnamefont
  {Ferrara}}, \bibinfo {author} {\bibfnamefont {A.~F.}\ \bibnamefont
  {Grillo}},\ and\ \bibinfo {author} {\bibfnamefont {R.}~\bibnamefont
  {Gatto}},\ }\bibfield  {title} {\bibinfo {title} {{Tensor representations of
  conformal algebra and conformally covariant operator product expansion}},\
  }\href {https://doi.org/10.1016/0003-4916(73)90446-6} {\bibfield  {journal}
  {\bibinfo  {journal} {Annals Phys.}\ }\textbf {\bibinfo {volume} {76}},\
  \bibinfo {pages} {161} (\bibinfo {year} {1973})}\BibitemShut {NoStop}%
\bibitem [{\citenamefont {Belavin}\ \emph {et~al.}(1984)\citenamefont
  {Belavin}, \citenamefont {Polyakov},\ and\ \citenamefont
  {Zamolodchikov}}]{Belavin:1984vu}%
  \BibitemOpen
  \bibfield  {author} {\bibinfo {author} {\bibfnamefont {A.~A.}\ \bibnamefont
  {Belavin}}, \bibinfo {author} {\bibfnamefont {A.~M.}\ \bibnamefont
  {Polyakov}},\ and\ \bibinfo {author} {\bibfnamefont {A.~B.}\ \bibnamefont
  {Zamolodchikov}},\ }\bibfield  {title} {\bibinfo {title} {{Infinite Conformal
  Symmetry in Two-Dimensional Quantum Field Theory}},\ }\href
  {https://doi.org/10.1016/0550-3213(84)90052-X} {\bibfield  {journal}
  {\bibinfo  {journal} {Nucl. Phys. B}\ }\textbf {\bibinfo {volume} {241}},\
  \bibinfo {pages} {333} (\bibinfo {year} {1984})}\BibitemShut {NoStop}%
\bibitem [{\citenamefont {Rattazzi}\ \emph {et~al.}(2008)\citenamefont
  {Rattazzi}, \citenamefont {Rychkov}, \citenamefont {Tonni},\ and\
  \citenamefont {Vichi}}]{Rattazzi:2008pe}%
  \BibitemOpen
  \bibfield  {author} {\bibinfo {author} {\bibfnamefont {R.}~\bibnamefont
  {Rattazzi}}, \bibinfo {author} {\bibfnamefont {V.~S.}\ \bibnamefont
  {Rychkov}}, \bibinfo {author} {\bibfnamefont {E.}~\bibnamefont {Tonni}},\
  and\ \bibinfo {author} {\bibfnamefont {A.}~\bibnamefont {Vichi}},\ }\bibfield
   {title} {\bibinfo {title} {{Bounding scalar operator dimensions in 4D
  CFT}},\ }\href {https://doi.org/10.1088/1126-6708/2008/12/031} {\bibfield
  {journal} {\bibinfo  {journal} {JHEP}\ }\textbf {\bibinfo {volume} {12}},\
  \bibinfo {pages} {031}},\ \Eprint {https://arxiv.org/abs/0807.0004}
  {arXiv:0807.0004 [hep-th]} \BibitemShut {NoStop}%
\bibitem [{\citenamefont {Poland}\ \emph {et~al.}(2019)\citenamefont {Poland},
  \citenamefont {Rychkov},\ and\ \citenamefont {Vichi}}]{Poland:2018epd}%
  \BibitemOpen
  \bibfield  {author} {\bibinfo {author} {\bibfnamefont {D.}~\bibnamefont
  {Poland}}, \bibinfo {author} {\bibfnamefont {S.}~\bibnamefont {Rychkov}},\
  and\ \bibinfo {author} {\bibfnamefont {A.}~\bibnamefont {Vichi}},\ }\bibfield
   {title} {\bibinfo {title} {{The Conformal Bootstrap: Theory, Numerical
  Techniques, and Applications}},\ }\href
  {https://doi.org/10.1103/RevModPhys.91.015002} {\bibfield  {journal}
  {\bibinfo  {journal} {Rev. Mod. Phys.}\ }\textbf {\bibinfo {volume} {91}},\
  \bibinfo {pages} {015002} (\bibinfo {year} {2019})},\ \Eprint
  {https://arxiv.org/abs/1805.04405} {arXiv:1805.04405 [hep-th]} \BibitemShut
  {NoStop}%
\bibitem [{\citenamefont {El-Showk}\ \emph {et~al.}(2012)\citenamefont
  {El-Showk}, \citenamefont {Paulos}, \citenamefont {Poland}, \citenamefont
  {Rychkov}, \citenamefont {Simmons-Duffin},\ and\ \citenamefont
  {Vichi}}]{El-Showk:2012cjh}%
  \BibitemOpen
  \bibfield  {author} {\bibinfo {author} {\bibfnamefont {S.}~\bibnamefont
  {El-Showk}}, \bibinfo {author} {\bibfnamefont {M.~F.}\ \bibnamefont
  {Paulos}}, \bibinfo {author} {\bibfnamefont {D.}~\bibnamefont {Poland}},
  \bibinfo {author} {\bibfnamefont {S.}~\bibnamefont {Rychkov}}, \bibinfo
  {author} {\bibfnamefont {D.}~\bibnamefont {Simmons-Duffin}},\ and\ \bibinfo
  {author} {\bibfnamefont {A.}~\bibnamefont {Vichi}},\ }\bibfield  {title}
  {\bibinfo {title} {{Solving the 3D Ising Model with the Conformal
  Bootstrap}},\ }\href {https://doi.org/10.1103/PhysRevD.86.025022} {\bibfield
  {journal} {\bibinfo  {journal} {Phys. Rev. D}\ }\textbf {\bibinfo {volume}
  {86}},\ \bibinfo {pages} {025022} (\bibinfo {year} {2012})},\ \Eprint
  {https://arxiv.org/abs/1203.6064} {arXiv:1203.6064 [hep-th]} \BibitemShut
  {NoStop}%
\bibitem [{\citenamefont {El-Showk}\ \emph {et~al.}(2014)\citenamefont
  {El-Showk}, \citenamefont {Paulos}, \citenamefont {Poland}, \citenamefont
  {Rychkov}, \citenamefont {Simmons-Duffin},\ and\ \citenamefont
  {Vichi}}]{El-Showk:2014dwa}%
  \BibitemOpen
  \bibfield  {author} {\bibinfo {author} {\bibfnamefont {S.}~\bibnamefont
  {El-Showk}}, \bibinfo {author} {\bibfnamefont {M.~F.}\ \bibnamefont
  {Paulos}}, \bibinfo {author} {\bibfnamefont {D.}~\bibnamefont {Poland}},
  \bibinfo {author} {\bibfnamefont {S.}~\bibnamefont {Rychkov}}, \bibinfo
  {author} {\bibfnamefont {D.}~\bibnamefont {Simmons-Duffin}},\ and\ \bibinfo
  {author} {\bibfnamefont {A.}~\bibnamefont {Vichi}},\ }\bibfield  {title}
  {\bibinfo {title} {{Solving the 3d Ising Model with the Conformal Bootstrap
  II. c-Minimization and Precise Critical Exponents}},\ }\href
  {https://doi.org/10.1007/s10955-014-1042-7} {\bibfield  {journal} {\bibinfo
  {journal} {J. Stat. Phys.}\ }\textbf {\bibinfo {volume} {157}},\ \bibinfo
  {pages} {869} (\bibinfo {year} {2014})},\ \Eprint
  {https://arxiv.org/abs/1403.4545} {arXiv:1403.4545 [hep-th]} \BibitemShut
  {NoStop}%
\bibitem [{\citenamefont {Kos}\ \emph {et~al.}(2014{\natexlab{a}})\citenamefont
  {Kos}, \citenamefont {Poland},\ and\ \citenamefont
  {Simmons-Duffin}}]{Kos:2014bka}%
  \BibitemOpen
  \bibfield  {author} {\bibinfo {author} {\bibfnamefont {F.}~\bibnamefont
  {Kos}}, \bibinfo {author} {\bibfnamefont {D.}~\bibnamefont {Poland}},\ and\
  \bibinfo {author} {\bibfnamefont {D.}~\bibnamefont {Simmons-Duffin}},\
  }\bibfield  {title} {\bibinfo {title} {{Bootstrapping Mixed Correlators in
  the 3D Ising Model}},\ }\href {https://doi.org/10.1007/JHEP11(2014)109}
  {\bibfield  {journal} {\bibinfo  {journal} {JHEP}\ }\textbf {\bibinfo
  {volume} {11}},\ \bibinfo {pages} {109}},\ \Eprint
  {https://arxiv.org/abs/1406.4858} {arXiv:1406.4858 [hep-th]} \BibitemShut
  {NoStop}%
\bibitem [{\citenamefont {Kos}\ \emph {et~al.}(2015)\citenamefont {Kos},
  \citenamefont {Poland}, \citenamefont {Simmons-Duffin},\ and\ \citenamefont
  {Vichi}}]{Kos:2015mba}%
  \BibitemOpen
  \bibfield  {author} {\bibinfo {author} {\bibfnamefont {F.}~\bibnamefont
  {Kos}}, \bibinfo {author} {\bibfnamefont {D.}~\bibnamefont {Poland}},
  \bibinfo {author} {\bibfnamefont {D.}~\bibnamefont {Simmons-Duffin}},\ and\
  \bibinfo {author} {\bibfnamefont {A.}~\bibnamefont {Vichi}},\ }\bibfield
  {title} {\bibinfo {title} {{Bootstrapping the O(N) Archipelago}},\ }\href
  {https://doi.org/10.1007/JHEP11(2015)106} {\bibfield  {journal} {\bibinfo
  {journal} {JHEP}\ }\textbf {\bibinfo {volume} {11}},\ \bibinfo {pages}
  {106}},\ \Eprint {https://arxiv.org/abs/1504.07997} {arXiv:1504.07997
  [hep-th]} \BibitemShut {NoStop}%
\bibitem [{\citenamefont {Poland}\ \emph {et~al.}(2012)\citenamefont {Poland},
  \citenamefont {Simmons-Duffin},\ and\ \citenamefont {Vichi}}]{Poland:2011ey}%
  \BibitemOpen
  \bibfield  {author} {\bibinfo {author} {\bibfnamefont {D.}~\bibnamefont
  {Poland}}, \bibinfo {author} {\bibfnamefont {D.}~\bibnamefont
  {Simmons-Duffin}},\ and\ \bibinfo {author} {\bibfnamefont {A.}~\bibnamefont
  {Vichi}},\ }\bibfield  {title} {\bibinfo {title} {{Carving Out the Space of
  4D CFTs}},\ }\href {https://doi.org/10.1007/JHEP05(2012)110} {\bibfield
  {journal} {\bibinfo  {journal} {JHEP}\ }\textbf {\bibinfo {volume} {05}},\
  \bibinfo {pages} {110}},\ \Eprint {https://arxiv.org/abs/1109.5176}
  {arXiv:1109.5176 [hep-th]} \BibitemShut {NoStop}%
\bibitem [{\citenamefont {Caracciolo}\ \emph {et~al.}(2014)\citenamefont
  {Caracciolo}, \citenamefont {Castedo~Echeverri}, \citenamefont {von
  Harling},\ and\ \citenamefont {Serone}}]{Caracciolo:2014cxa}%
  \BibitemOpen
  \bibfield  {author} {\bibinfo {author} {\bibfnamefont {F.}~\bibnamefont
  {Caracciolo}}, \bibinfo {author} {\bibfnamefont {A.}~\bibnamefont
  {Castedo~Echeverri}}, \bibinfo {author} {\bibfnamefont {B.}~\bibnamefont {von
  Harling}},\ and\ \bibinfo {author} {\bibfnamefont {M.}~\bibnamefont
  {Serone}},\ }\bibfield  {title} {\bibinfo {title} {{Bounds on OPE
  Coefficients in 4D Conformal Field Theories}},\ }\href
  {https://doi.org/10.1007/JHEP10(2014)020} {\bibfield  {journal} {\bibinfo
  {journal} {JHEP}\ }\textbf {\bibinfo {volume} {10}},\ \bibinfo {pages}
  {020}},\ \Eprint {https://arxiv.org/abs/1406.7845} {arXiv:1406.7845 [hep-th]}
  \BibitemShut {NoStop}%
\bibitem [{\citenamefont {Nakayama}(2016)}]{Nakayama:2016knq}%
  \BibitemOpen
  \bibfield  {author} {\bibinfo {author} {\bibfnamefont {Y.}~\bibnamefont
  {Nakayama}},\ }\bibfield  {title} {\bibinfo {title} {{Bootstrap bound for
  conformal multi-flavor QCD on lattice}},\ }\href
  {https://doi.org/10.1007/JHEP07(2016)038} {\bibfield  {journal} {\bibinfo
  {journal} {JHEP}\ }\textbf {\bibinfo {volume} {07}},\ \bibinfo {pages}
  {038}},\ \Eprint {https://arxiv.org/abs/1605.04052} {arXiv:1605.04052
  [hep-th]} \BibitemShut {NoStop}%
\bibitem [{\citenamefont {Nakayama}(2018)}]{Nakayama:2017vdd}%
  \BibitemOpen
  \bibfield  {author} {\bibinfo {author} {\bibfnamefont {Y.}~\bibnamefont
  {Nakayama}},\ }\bibfield  {title} {\bibinfo {title} {{Bootstrap experiments
  on higher dimensional CFTs}},\ }\href
  {https://doi.org/10.1142/S0217751X18500367} {\bibfield  {journal} {\bibinfo
  {journal} {Int. J. Mod. Phys. A}\ }\textbf {\bibinfo {volume} {33}},\
  \bibinfo {pages} {1850036} (\bibinfo {year} {2018})},\ \Eprint
  {https://arxiv.org/abs/1705.02744} {arXiv:1705.02744 [hep-th]} \BibitemShut
  {NoStop}%
\bibitem [{\citenamefont {Li}(2018)}]{Li:2018lyb}%
  \BibitemOpen
  \bibfield  {author} {\bibinfo {author} {\bibfnamefont {Z.}~\bibnamefont
  {Li}},\ }\href@noop {} {\bibinfo {title} {{Bootstrapping conformal QED$_3$
  and deconfined quantum critical point}}} (\bibinfo {year} {2018}),\ \Eprint
  {https://arxiv.org/abs/1812.09281} {arXiv:1812.09281 [hep-th]} \BibitemShut
  {NoStop}%
\bibitem [{\citenamefont {Stergiou}(2018)}]{Stergiou:2018gjj}%
  \BibitemOpen
  \bibfield  {author} {\bibinfo {author} {\bibfnamefont {A.}~\bibnamefont
  {Stergiou}},\ }\bibfield  {title} {\bibinfo {title} {{Bootstrapping
  hypercubic and hypertetrahedral theories in three dimensions}},\ }\href
  {https://doi.org/10.1007/JHEP05(2018)035} {\bibfield  {journal} {\bibinfo
  {journal} {JHEP}\ }\textbf {\bibinfo {volume} {05}},\ \bibinfo {pages}
  {035}},\ \Eprint {https://arxiv.org/abs/1801.07127} {arXiv:1801.07127
  [hep-th]} \BibitemShut {NoStop}%
\bibitem [{\citenamefont {Stergiou}(2019)}]{Stergiou:2019dcv}%
  \BibitemOpen
  \bibfield  {author} {\bibinfo {author} {\bibfnamefont {A.}~\bibnamefont
  {Stergiou}},\ }\bibfield  {title} {\bibinfo {title} {{Bootstrapping MN and
  Tetragonal CFTs in Three Dimensions}},\ }\href
  {https://doi.org/10.21468/SciPostPhys.7.1.010} {\bibfield  {journal}
  {\bibinfo  {journal} {SciPost Phys.}\ }\textbf {\bibinfo {volume} {7}},\
  \bibinfo {pages} {010} (\bibinfo {year} {2019})},\ \Eprint
  {https://arxiv.org/abs/1904.00017} {arXiv:1904.00017 [hep-th]} \BibitemShut
  {NoStop}%
\bibitem [{\citenamefont {Li}\ and\ \citenamefont {Poland}(2021)}]{Li:2020bnb}%
  \BibitemOpen
  \bibfield  {author} {\bibinfo {author} {\bibfnamefont {Z.}~\bibnamefont
  {Li}}\ and\ \bibinfo {author} {\bibfnamefont {D.}~\bibnamefont {Poland}},\
  }\bibfield  {title} {\bibinfo {title} {{Searching for gauge theories with the
  conformal bootstrap}},\ }\href {https://doi.org/10.1007/JHEP03(2021)172}
  {\bibfield  {journal} {\bibinfo  {journal} {JHEP}\ }\textbf {\bibinfo
  {volume} {03}},\ \bibinfo {pages} {172}},\ \Eprint
  {https://arxiv.org/abs/2005.01721} {arXiv:2005.01721 [hep-th]} \BibitemShut
  {NoStop}%
\bibitem [{\citenamefont {Rattazzi}\ \emph {et~al.}(2011)\citenamefont
  {Rattazzi}, \citenamefont {Rychkov},\ and\ \citenamefont
  {Vichi}}]{Rattazzi:2010yc}%
  \BibitemOpen
  \bibfield  {author} {\bibinfo {author} {\bibfnamefont {R.}~\bibnamefont
  {Rattazzi}}, \bibinfo {author} {\bibfnamefont {S.}~\bibnamefont {Rychkov}},\
  and\ \bibinfo {author} {\bibfnamefont {A.}~\bibnamefont {Vichi}},\ }\bibfield
   {title} {\bibinfo {title} {{Bounds in 4D Conformal Field Theories with
  Global Symmetry}},\ }\href {https://doi.org/10.1088/1751-8113/44/3/035402}
  {\bibfield  {journal} {\bibinfo  {journal} {J. Phys. A}\ }\textbf {\bibinfo
  {volume} {44}},\ \bibinfo {pages} {035402} (\bibinfo {year} {2011})},\
  \Eprint {https://arxiv.org/abs/1009.5985} {arXiv:1009.5985 [hep-th]}
  \BibitemShut {NoStop}%
\bibitem [{\citenamefont {Go}\ and\ \citenamefont
  {Tachikawa}(2019)}]{Go:2019lke}%
  \BibitemOpen
  \bibfield  {author} {\bibinfo {author} {\bibfnamefont {M.}~\bibnamefont
  {Go}}\ and\ \bibinfo {author} {\bibfnamefont {Y.}~\bibnamefont {Tachikawa}},\
  }\bibfield  {title} {\bibinfo {title} {{autoboot: A generator of bootstrap
  equations with global symmetry}},\ }\href
  {https://doi.org/10.1007/JHEP06(2019)084} {\bibfield  {journal} {\bibinfo
  {journal} {JHEP}\ }\textbf {\bibinfo {volume} {06}},\ \bibinfo {pages}
  {084}},\ \Eprint {https://arxiv.org/abs/1903.10522} {arXiv:1903.10522
  [hep-th]} \BibitemShut {NoStop}%
\bibitem [{\citenamefont {Dolan}\ and\ \citenamefont
  {Osborn}(2001)}]{Dolan:2000ut}%
  \BibitemOpen
  \bibfield  {author} {\bibinfo {author} {\bibfnamefont {F.~A.}\ \bibnamefont
  {Dolan}}\ and\ \bibinfo {author} {\bibfnamefont {H.}~\bibnamefont {Osborn}},\
  }\bibfield  {title} {\bibinfo {title} {{Conformal four point functions and
  the operator product expansion}},\ }\href
  {https://doi.org/10.1016/S0550-3213(01)00013-X} {\bibfield  {journal}
  {\bibinfo  {journal} {Nucl. Phys. B}\ }\textbf {\bibinfo {volume} {599}},\
  \bibinfo {pages} {459} (\bibinfo {year} {2001})},\ \Eprint
  {https://arxiv.org/abs/hep-th/0011040} {arXiv:hep-th/0011040} \BibitemShut
  {NoStop}%
\bibitem [{\citenamefont {Dolan}\ and\ \citenamefont
  {Osborn}(2011)}]{Dolan:2011dv}%
  \BibitemOpen
  \bibfield  {author} {\bibinfo {author} {\bibfnamefont {F.~A.}\ \bibnamefont
  {Dolan}}\ and\ \bibinfo {author} {\bibfnamefont {H.}~\bibnamefont {Osborn}},\
  }\href@noop {} {\bibinfo {title} {{Conformal Partial Waves: Further
  Mathematical Results}}} (\bibinfo {year} {2011}),\ \Eprint
  {https://arxiv.org/abs/1108.6194} {arXiv:1108.6194 [hep-th]} \BibitemShut
  {NoStop}%
\bibitem [{\citenamefont {Dolan}\ and\ \citenamefont
  {Osborn}(2004)}]{Dolan:2003hv}%
  \BibitemOpen
  \bibfield  {author} {\bibinfo {author} {\bibfnamefont {F.~A.}\ \bibnamefont
  {Dolan}}\ and\ \bibinfo {author} {\bibfnamefont {H.}~\bibnamefont {Osborn}},\
  }\bibfield  {title} {\bibinfo {title} {{Conformal partial waves and the
  operator product expansion}},\ }\href
  {https://doi.org/10.1016/j.nuclphysb.2003.11.016} {\bibfield  {journal}
  {\bibinfo  {journal} {Nucl. Phys. B}\ }\textbf {\bibinfo {volume} {678}},\
  \bibinfo {pages} {491} (\bibinfo {year} {2004})},\ \Eprint
  {https://arxiv.org/abs/hep-th/0309180} {arXiv:hep-th/0309180} \BibitemShut
  {NoStop}%
\bibitem [{\citenamefont {Gadde}(2017)}]{Gadde:2017sjg}%
  \BibitemOpen
  \bibfield  {author} {\bibinfo {author} {\bibfnamefont {A.}~\bibnamefont
  {Gadde}},\ }\href@noop {} {\bibinfo {title} {{In search of conformal
  theories}}} (\bibinfo {year} {2017}),\ \Eprint
  {https://arxiv.org/abs/1702.07362} {arXiv:1702.07362 [hep-th]} \BibitemShut
  {NoStop}%
\bibitem [{\citenamefont {Liu}\ \emph {et~al.}(2019)\citenamefont {Liu},
  \citenamefont {Perlmutter}, \citenamefont {Rosenhaus},\ and\ \citenamefont
  {Simmons-Duffin}}]{Liu:2018jhs}%
  \BibitemOpen
  \bibfield  {author} {\bibinfo {author} {\bibfnamefont {J.}~\bibnamefont
  {Liu}}, \bibinfo {author} {\bibfnamefont {E.}~\bibnamefont {Perlmutter}},
  \bibinfo {author} {\bibfnamefont {V.}~\bibnamefont {Rosenhaus}},\ and\
  \bibinfo {author} {\bibfnamefont {D.}~\bibnamefont {Simmons-Duffin}},\
  }\bibfield  {title} {\bibinfo {title} {{$d$-dimensional SYK, AdS Loops, and
  $6j$ Symbols}},\ }\href {https://doi.org/10.1007/JHEP03(2019)052} {\bibfield
  {journal} {\bibinfo  {journal} {JHEP}\ }\textbf {\bibinfo {volume} {03}},\
  \bibinfo {pages} {052}},\ \Eprint {https://arxiv.org/abs/1808.00612}
  {arXiv:1808.00612 [hep-th]} \BibitemShut {NoStop}%
\bibitem [{\citenamefont {Rychkov}(2017)}]{Rychkov:2016iqz}%
  \BibitemOpen
  \bibfield  {author} {\bibinfo {author} {\bibfnamefont {S.}~\bibnamefont
  {Rychkov}},\ }\bibfield  {title} {\bibinfo {title} {Epfl lectures on
  conformal field theory in d \ensuremath{>}= 3 dimensions},\ }\bibfield
  {journal} {\bibinfo  {journal} {SpringerBriefs in Physics}\ }\href
  {https://doi.org/10.1007/978-3-319-43626-5} {10.1007/978-3-319-43626-5}
  (\bibinfo {year} {2017})\BibitemShut {NoStop}%
\bibitem [{\citenamefont {Simmons-Duffin}(2016)}]{Simmons-Duffin:2016gjk}%
  \BibitemOpen
  \bibfield  {author} {\bibinfo {author} {\bibfnamefont {D.}~\bibnamefont
  {Simmons-Duffin}},\ }\href@noop {} {\bibinfo {title} {Tasi lectures on the
  conformal bootstrap}} (\bibinfo {year} {2016}),\ \Eprint
  {https://arxiv.org/abs/1602.07982} {arXiv:1602.07982 [hep-th]} \BibitemShut
  {NoStop}%
\bibitem [{\citenamefont {Chester}(2019)}]{Chester:2019wfx}%
  \BibitemOpen
  \bibfield  {author} {\bibinfo {author} {\bibfnamefont {S.~M.}\ \bibnamefont
  {Chester}},\ }\href@noop {} {\bibinfo {title} {{Weizmann Lectures on the
  Numerical Conformal Bootstrap}}} (\bibinfo {year} {2019}),\ \Eprint
  {https://arxiv.org/abs/1907.05147} {arXiv:1907.05147 [hep-th]} \BibitemShut
  {NoStop}%
\bibitem [{\citenamefont {Berkooz}\ \emph {et~al.}(2014)\citenamefont
  {Berkooz}, \citenamefont {Yacoby},\ and\ \citenamefont
  {Zait}}]{Berkooz:2014yda}%
  \BibitemOpen
  \bibfield  {author} {\bibinfo {author} {\bibfnamefont {M.}~\bibnamefont
  {Berkooz}}, \bibinfo {author} {\bibfnamefont {R.}~\bibnamefont {Yacoby}},\
  and\ \bibinfo {author} {\bibfnamefont {A.}~\bibnamefont {Zait}},\ }\bibfield
  {title} {\bibinfo {title} {{Bounds on $\mathcal{N} = 1$ superconformal
  theories with global symmetries}},\ }\href
  {https://doi.org/10.1007/JHEP01(2015)132} {\bibfield  {journal} {\bibinfo
  {journal} {JHEP}\ }\textbf {\bibinfo {volume} {08}},\ \bibinfo {pages}
  {008}},\ \bibinfo {note} {[Erratum: JHEP 01, 132 (2015)]},\ \Eprint
  {https://arxiv.org/abs/1402.6068} {arXiv:1402.6068 [hep-th]} \BibitemShut
  {NoStop}%
\bibitem [{\citenamefont {{Kawamura}}(1998)}]{1998JPCM10}%
  \BibitemOpen
  \bibfield  {author} {\bibinfo {author} {\bibfnamefont {H.}~\bibnamefont
  {{Kawamura}}},\ }\bibfield  {title} {\bibinfo {title} {{REVIEW ARTICLE:
  Universality of phase transitions of frustrated antiferromagnets}},\ }\href
  {https://doi.org/10.1088/0953-8984/10/22/004} {\bibfield  {journal} {\bibinfo
   {journal} {Journal of Physics Condensed Matter}\ }\textbf {\bibinfo {volume}
  {10}},\ \bibinfo {pages} {4707} (\bibinfo {year} {1998})},\ \Eprint
  {https://arxiv.org/abs/cond-mat/9805134} {arXiv:cond-mat/9805134
  [cond-mat.stat-mech]} \BibitemShut {NoStop}%
\bibitem [{\citenamefont {Delamotte}\ \emph {et~al.}(2004)\citenamefont
  {Delamotte}, \citenamefont {Mouhanna},\ and\ \citenamefont
  {Tissier}}]{Delamotte:2003dw}%
  \BibitemOpen
  \bibfield  {author} {\bibinfo {author} {\bibfnamefont {B.}~\bibnamefont
  {Delamotte}}, \bibinfo {author} {\bibfnamefont {D.}~\bibnamefont
  {Mouhanna}},\ and\ \bibinfo {author} {\bibfnamefont {M.}~\bibnamefont
  {Tissier}},\ }\bibfield  {title} {\bibinfo {title} {{Nonperturbative
  renormalization group approach to frustrated magnets}},\ }\href
  {https://doi.org/10.1103/PhysRevB.69.134413} {\bibfield  {journal} {\bibinfo
  {journal} {Phys. Rev. B}\ }\textbf {\bibinfo {volume} {69}},\ \bibinfo
  {pages} {134413} (\bibinfo {year} {2004})},\ \Eprint
  {https://arxiv.org/abs/cond-mat/0309101} {arXiv:cond-mat/0309101}
  \BibitemShut {NoStop}%
\bibitem [{\citenamefont {Nakayama}\ and\ \citenamefont
  {Ohtsuki}(2014)}]{Nakayama:2014lva}%
  \BibitemOpen
  \bibfield  {author} {\bibinfo {author} {\bibfnamefont {Y.}~\bibnamefont
  {Nakayama}}\ and\ \bibinfo {author} {\bibfnamefont {T.}~\bibnamefont
  {Ohtsuki}},\ }\bibfield  {title} {\bibinfo {title} {{Approaching the
  conformal window of $O(n)\times O(m)$ symmetric Landau-Ginzburg models using
  the conformal bootstrap}},\ }\href
  {https://doi.org/10.1103/PhysRevD.89.126009} {\bibfield  {journal} {\bibinfo
  {journal} {Phys. Rev. D}\ }\textbf {\bibinfo {volume} {89}},\ \bibinfo
  {pages} {126009} (\bibinfo {year} {2014})},\ \Eprint
  {https://arxiv.org/abs/1404.0489} {arXiv:1404.0489 [hep-th]} \BibitemShut
  {NoStop}%
\bibitem [{\citenamefont {Iha}\ \emph {et~al.}(2016)\citenamefont {Iha},
  \citenamefont {Makino},\ and\ \citenamefont {Suzuki}}]{Iha:2016ppj}%
  \BibitemOpen
  \bibfield  {author} {\bibinfo {author} {\bibfnamefont {H.}~\bibnamefont
  {Iha}}, \bibinfo {author} {\bibfnamefont {H.}~\bibnamefont {Makino}},\ and\
  \bibinfo {author} {\bibfnamefont {H.}~\bibnamefont {Suzuki}},\ }\bibfield
  {title} {\bibinfo {title} {{Upper bound on the mass anomalous dimension in
  many-flavor gauge theories: a conformal bootstrap approach}},\ }\href
  {https://doi.org/10.1093/ptep/ptw046} {\bibfield  {journal} {\bibinfo
  {journal} {PTEP}\ }\textbf {\bibinfo {volume} {2016}},\ \bibinfo {pages}
  {053B03} (\bibinfo {year} {2016})},\ \Eprint
  {https://arxiv.org/abs/1603.01995} {arXiv:1603.01995 [hep-th]} \BibitemShut
  {NoStop}%
\bibitem [{\citenamefont {Wess}\ and\ \citenamefont
  {Zumino}(1971)}]{Wess:1971yu}%
  \BibitemOpen
  \bibfield  {author} {\bibinfo {author} {\bibfnamefont {J.}~\bibnamefont
  {Wess}}\ and\ \bibinfo {author} {\bibfnamefont {B.}~\bibnamefont {Zumino}},\
  }\bibfield  {title} {\bibinfo {title} {{Consequences of anomalous Ward
  identities}},\ }\href {https://doi.org/10.1016/0370-2693(71)90582-X}
  {\bibfield  {journal} {\bibinfo  {journal} {Phys. Lett. B}\ }\textbf
  {\bibinfo {volume} {37}},\ \bibinfo {pages} {95} (\bibinfo {year}
  {1971})}\BibitemShut {NoStop}%
\bibitem [{\citenamefont {Witten}(1983)}]{Witten:1983tw}%
  \BibitemOpen
  \bibfield  {author} {\bibinfo {author} {\bibfnamefont {E.}~\bibnamefont
  {Witten}},\ }\bibfield  {title} {\bibinfo {title} {{Global Aspects of Current
  Algebra}},\ }\href {https://doi.org/10.1016/0550-3213(83)90063-9} {\bibfield
  {journal} {\bibinfo  {journal} {Nucl. Phys. B}\ }\textbf {\bibinfo {volume}
  {223}},\ \bibinfo {pages} {422} (\bibinfo {year} {1983})}\BibitemShut
  {NoStop}%
\bibitem [{\citenamefont {Knizhnik}\ and\ \citenamefont
  {Zamolodchikov}(1984)}]{Knizhnik:1984nr}%
  \BibitemOpen
  \bibfield  {author} {\bibinfo {author} {\bibfnamefont {V.~G.}\ \bibnamefont
  {Knizhnik}}\ and\ \bibinfo {author} {\bibfnamefont {A.~B.}\ \bibnamefont
  {Zamolodchikov}},\ }\bibfield  {title} {\bibinfo {title} {{Current Algebra
  and Wess-Zumino Model in Two-Dimensions}},\ }\href
  {https://doi.org/10.1016/0550-3213(84)90374-2} {\bibfield  {journal}
  {\bibinfo  {journal} {Nucl. Phys. B}\ }\textbf {\bibinfo {volume} {247}},\
  \bibinfo {pages} {83} (\bibinfo {year} {1984})}\BibitemShut {NoStop}%
\bibitem [{\citenamefont {Kos}\ \emph {et~al.}(2014{\natexlab{b}})\citenamefont
  {Kos}, \citenamefont {Poland},\ and\ \citenamefont
  {Simmons-Duffin}}]{Kos:2013tga}%
  \BibitemOpen
  \bibfield  {author} {\bibinfo {author} {\bibfnamefont {F.}~\bibnamefont
  {Kos}}, \bibinfo {author} {\bibfnamefont {D.}~\bibnamefont {Poland}},\ and\
  \bibinfo {author} {\bibfnamefont {D.}~\bibnamefont {Simmons-Duffin}},\
  }\bibfield  {title} {\bibinfo {title} {{Bootstrapping the $O(N)$ vector
  models}},\ }\href {https://doi.org/10.1007/JHEP06(2014)091} {\bibfield
  {journal} {\bibinfo  {journal} {JHEP}\ }\textbf {\bibinfo {volume} {06}},\
  \bibinfo {pages} {091}},\ \Eprint {https://arxiv.org/abs/1307.6856}
  {arXiv:1307.6856 [hep-th]} \BibitemShut {NoStop}%
\bibitem [{\citenamefont {Appelquist}\ \emph {et~al.}(1988)\citenamefont
  {Appelquist}, \citenamefont {Nash},\ and\ \citenamefont
  {Wijewardhana}}]{Appelquist:1988sr}%
  \BibitemOpen
  \bibfield  {author} {\bibinfo {author} {\bibfnamefont {T.}~\bibnamefont
  {Appelquist}}, \bibinfo {author} {\bibfnamefont {D.}~\bibnamefont {Nash}},\
  and\ \bibinfo {author} {\bibfnamefont {L.~C.~R.}\ \bibnamefont
  {Wijewardhana}},\ }\bibfield  {title} {\bibinfo {title} {{Critical Behavior
  in (2+1)-Dimensional QED}},\ }\href
  {https://doi.org/10.1103/PhysRevLett.60.2575} {\bibfield  {journal} {\bibinfo
   {journal} {Phys. Rev. Lett.}\ }\textbf {\bibinfo {volume} {60}},\ \bibinfo
  {pages} {2575} (\bibinfo {year} {1988})}\BibitemShut {NoStop}%
\bibitem [{\citenamefont {Nash}(1989)}]{Nash:1989xx}%
  \BibitemOpen
  \bibfield  {author} {\bibinfo {author} {\bibfnamefont {D.}~\bibnamefont
  {Nash}},\ }\bibfield  {title} {\bibinfo {title} {{Higher Order Corrections in
  (2+1)-Dimensional QED}},\ }\href
  {https://doi.org/10.1103/PhysRevLett.62.3024} {\bibfield  {journal} {\bibinfo
   {journal} {Phys. Rev. Lett.}\ }\textbf {\bibinfo {volume} {62}},\ \bibinfo
  {pages} {3024} (\bibinfo {year} {1989})}\BibitemShut {NoStop}%
\bibitem [{\citenamefont {Caswell}(1974)}]{Caswell:1974gg}%
  \BibitemOpen
  \bibfield  {author} {\bibinfo {author} {\bibfnamefont {W.~E.}\ \bibnamefont
  {Caswell}},\ }\bibfield  {title} {\bibinfo {title} {{Asymptotic Behavior of
  Nonabelian Gauge Theories to Two Loop Order}},\ }\href
  {https://doi.org/10.1103/PhysRevLett.33.244} {\bibfield  {journal} {\bibinfo
  {journal} {Phys. Rev. Lett.}\ }\textbf {\bibinfo {volume} {33}},\ \bibinfo
  {pages} {244} (\bibinfo {year} {1974})}\BibitemShut {NoStop}%
\bibitem [{\citenamefont {Banks}\ and\ \citenamefont
  {Zaks}(1982)}]{Banks:1981nn}%
  \BibitemOpen
  \bibfield  {author} {\bibinfo {author} {\bibfnamefont {T.}~\bibnamefont
  {Banks}}\ and\ \bibinfo {author} {\bibfnamefont {A.}~\bibnamefont {Zaks}},\
  }\bibfield  {title} {\bibinfo {title} {{On the Phase Structure of Vector-Like
  Gauge Theories with Massless Fermions}},\ }\href
  {https://doi.org/10.1016/0550-3213(82)90035-9} {\bibfield  {journal}
  {\bibinfo  {journal} {Nucl. Phys. B}\ }\textbf {\bibinfo {volume} {196}},\
  \bibinfo {pages} {189} (\bibinfo {year} {1982})}\BibitemShut {NoStop}%
\bibitem [{\citenamefont {Witten}(1984)}]{Witten:1983ar}%
  \BibitemOpen
  \bibfield  {author} {\bibinfo {author} {\bibfnamefont {E.}~\bibnamefont
  {Witten}},\ }\bibfield  {title} {\bibinfo {title} {{Nonabelian Bosonization
  in Two-Dimensions}},\ }\href {https://doi.org/10.1007/BF01215276} {\bibfield
  {journal} {\bibinfo  {journal} {Commun. Math. Phys.}\ }\textbf {\bibinfo
  {volume} {92}},\ \bibinfo {pages} {455} (\bibinfo {year} {1984})}\BibitemShut
  {NoStop}%
\bibitem [{\citenamefont {Gepner}(1985)}]{Gepner:1984au}%
  \BibitemOpen
  \bibfield  {author} {\bibinfo {author} {\bibfnamefont {D.}~\bibnamefont
  {Gepner}},\ }\bibfield  {title} {\bibinfo {title} {{Nonabelian Bosonization
  and Multiflavor {QED} and {QCD} in Two-dimensions}},\ }\href
  {https://doi.org/10.1016/0550-3213(85)90458-4} {\bibfield  {journal}
  {\bibinfo  {journal} {Nucl. Phys. B}\ }\textbf {\bibinfo {volume} {252}},\
  \bibinfo {pages} {481} (\bibinfo {year} {1985})}\BibitemShut {NoStop}%
\bibitem [{\citenamefont {Osborn}\ and\ \citenamefont
  {Stergiou}(2018)}]{Osborn:2017ucf}%
  \BibitemOpen
  \bibfield  {author} {\bibinfo {author} {\bibfnamefont {H.}~\bibnamefont
  {Osborn}}\ and\ \bibinfo {author} {\bibfnamefont {A.}~\bibnamefont
  {Stergiou}},\ }\bibfield  {title} {\bibinfo {title} {{Seeking fixed points in
  multiple coupling scalar theories in the $\epsilon$ expansion}},\ }\href
  {https://doi.org/10.1007/JHEP05(2018)051} {\bibfield  {journal} {\bibinfo
  {journal} {JHEP}\ }\textbf {\bibinfo {volume} {05}},\ \bibinfo {pages}
  {051}},\ \Eprint {https://arxiv.org/abs/1707.06165} {arXiv:1707.06165
  [hep-th]} \BibitemShut {NoStop}%
\bibitem [{\citenamefont {Rychkov}\ and\ \citenamefont
  {Stergiou}(2019)}]{Rychkov:2018vya}%
  \BibitemOpen
  \bibfield  {author} {\bibinfo {author} {\bibfnamefont {S.}~\bibnamefont
  {Rychkov}}\ and\ \bibinfo {author} {\bibfnamefont {A.}~\bibnamefont
  {Stergiou}},\ }\bibfield  {title} {\bibinfo {title} {{General Properties of
  Multiscalar RG Flows in $d=4-\varepsilon$}},\ }\href
  {https://doi.org/10.21468/SciPostPhys.6.1.008} {\bibfield  {journal}
  {\bibinfo  {journal} {SciPost Phys.}\ }\textbf {\bibinfo {volume} {6}},\
  \bibinfo {pages} {008} (\bibinfo {year} {2019})},\ \Eprint
  {https://arxiv.org/abs/1810.10541} {arXiv:1810.10541 [hep-th]} \BibitemShut
  {NoStop}%
\end{thebibliography}%

\end{document}